\documentclass[12pt]{article}
\pdfoutput=1

\usepackage{amsmath}
\usepackage{graphicx}
\usepackage{wrapfig}
\usepackage{amsfonts}
\usepackage{xcolor}
\usepackage{mathrsfs}

\input{epsf.sty}
\definecolor{brown}{cmyk}{0,1,.9,.2}

\def\timesbox{\hbox{$\scriptscriptstyle\times$}}
\def\ant{ {{\lower 1ex  \timesbox} \atop {\raise 1.5ex  \timesbox}}}
\newcommand\ZZ{{\hbox{ Z\kern-1.6mm Z}}}
\newcommand\RR{{\hbox{ R\kern-2.5mm R}}}
\newcommand{\Iop}{\relax{\rm I\kern-.18em I}}
\newcommand{\Lop}{\relax{\rm I\kern-.18em L}}
\newcommand{\dop}{\relax{\rm I\kern-.8em d}}
\newcommand{\one}{{\hbox{ 1\kern-1.2mm l}}}

\newcommand{\beq}{\begin{equation}}
\newcommand{\eeq}{\end{equation}}
\newcommand{\bea}{\begin{eqnarray}}
\newcommand{\eea}{\end{eqnarray}}

\newcommand{\lt}{\left}
\newcommand{\rt}{\right}

\newcommand{\del}{\partial}

\newcommand{\al}{\alpha}

\newcommand{\dlt}{\delta}

\newcommand{\omg}{\omega}

\newcommand{\Dlt}{\Delta}

\newcommand{\Omg}{\Omega}

\newcommand{\rb}{ r}
\newcommand{\zb}{ z}

\newcommand{\xib}{ \xi}

\newcommand{\eh}{\hat e}

\newcommand{\yh}{\hat y}

\newcommand{\cF}{{\cal F}}

\newcommand{\cM}{{\cal M}}

\newcommand{\ha}{\hbox{a}}
\newcommand{\hb}{\hbox{b}}

\newcommand{\hr}{\hbox{r}}
\newcommand{\hv}{\hbox{v}}

\newcommand{\Exp}{{\hbox{Exp}}}

\begin{document}

{}~
{}~

\vskip 2cm

\centerline{\Large \bf  General Construction of Tubular Geometry} 

\medskip

\vspace*{4.0ex}

\centerline{\large \rm Partha Mukhopadhyay }

\vspace*{4.0ex}

\centerline{\it The Institute of Mathematical Sciences}
\centerline{\it C.I.T. Campus, Taramani}
\centerline{\it Chennai 600113, India}

\centerline{and}

\centerline{\it Department of Physics and Astronomy}
\centerline{\it University of Kentucky, Lexington, KY-40506, USA}

\medskip

\centerline{E-mail: parthamu@imsc.res.in}

\vspace*{5.0ex}

\centerline{\bf Abstract}
\bigskip

We consider the problem of locally describing tubular geometry around a submanifold embedded in a (pseudo)Riemannian manifold in its general form. Given the geometry of ambient space in an arbitrary coordinate system and equations determining the submanifold in the same system, we compute the tubular expansion coefficients in terms of this {\it a priori data}. This is done by using an indirect method that crucially applies the tubular expansion theorem for vielbein previously derived. With an explicit construction involving the relevant coordinate and non-coordinate frames we verify consistency of the whole method up to quadratic order in vielbein expansion. Furthermore, we perform certain (long and tedious) higher order computation which verifies the first non-trivial spin connection term in the expansion for the first time. Earlier a similar method was used to compute tubular geometry in loop space. We explain this work in the light of our general construction.

\newpage

\tableofcontents

\baselineskip=18pt

\section{Motivation and summary}
\label{motiv}

In the context of (pseudo)Riemannian geometry, Fermi normal coordinate (FNC) \cite{FNC, FS} expansion, or in short tubular expansion \cite{tubes} has many physical applications. In most of the situations one uses such an expansion around a particle trajectory in curved space to study its local physics \cite{cl-particle, living-rev, cm-exact, kc-exact, bini-exact, klein-exact, qnt-particle}. Some of the higher dimensional situations are as follows \footnote{See \cite{const-qnt} for its use in ``constrained quantum mechanics".}. 

It was pointed out in \cite{semi-classical, tubular, cut-off} that use of tubular expansion may possibly be found in the context of generally covariant description of bound configurations. The general idea is as follows. The long distance behaviour is described as a particle degree of freedom (DOF) moving in the background geometry (target space). This is a subsector of the full set of DOF that is identified as the {\it slow} one in a generalized Born-Oppenheimer sense\footnote{The classical non-relativistic analogue of this is the centre of mass which does not have a straightforward generalization to the relativistic description.}. Then the target space must be sitting as a submanifold within the full configuration space. The higher derivative corrections to the long distance behaviour due to the finite size are to be computed from the tubular expansion around this submanifold. It is not clear at this stage if this description is generally applicable. However, the analysis of non-linear sigma model in terms of loop space quantum mechanics, as pursued in \cite{semi-classical}, is an explicit realization of this idea \footnote{In order to investigate whether any quantum bound configuration should have a description in such a tubular geometric framework, one may first try to construct a fully covariant string bits model and study the consistency requirements following the same principles of first quantized string theory. This is a work in progress \cite{csb}. If successful, one may then try to see if a string inspired general framework of this sort can be formulated independent of dynamics.}. 

Another interesting application may possibly be found in context of the recently discussed large-$D$ black holes \cite{emparan, minwalla}. The idea that seems to emerge from this work is that the solution outside of a black hole with a certain asymptotic geometry in large dimension ($D$) is given by a co-dimension one submanifold (membrane) representing the horizon in the same geometry. Moreover, the ${1\over D}$ corrections are supported in the tubular neighborhood of this submanifold. One may wonder if the construction can be rephrased with submanifold geometry more manifest which may help generalizing it to black holes in other backgrounds.

To avoid too much of digression at this point, another (technical) motivation has been discussed at the end in \S \ref{end}.

In any given situation, the usual approach for computing the desired expansion has so far been to construct FNC order-by-order and then compute the covariant expansion by directly applying the required coordinate transformation. Another approach that has been studied in the literature is to consider special submanifolds in certain specific ambient geometries so that FNC can be constructed, and consiquently tubular geometry can be evaluated, exactly \cite{cm-exact, kc-exact, bini-exact, klein-exact}.  

In this work we address the following most general form of the problem, which we hope will be fitting, for example, for the aforementioned black hole problem. We start with the geometry of ambient space given in an arbitrary {\it a priori} coordinate system and equations determining the submanifold in the same system. The problem is to compute the tubular geometry to arbitrary high orders in terms of these {\it a priori data}. In \cite{cut-off}, a similar problem was considered for the specific case of submanifold of vanishing loops sitting in the free loop space of a (pseudo)Riemannian manifold. The problem was solved by using an indirect method where the tubular expansion theorem of \cite{tubular} was crucially used. This theorem spells out all the tubular expansion coefficients of vielbein for an arbitrary submanifold embedding. Here we construct the most general version of the same argument.

Our final result can be qualitatively summarized as follows. We denote the a priori system describing the local geometry of the ambient space $L$ as,
\bea
z^{ a} ~, \quad ( a = 1, 2, \cdots , \dim L)~.
\label{zbar}
\eea
Equations determining the submanifold $M$ in the same system are given by,
\bea
z^{a} &=& f^{ a} (x) ~, 
\label{submanifold-eq}
\eea
where $x^{\al}$ $(\al = 1, 2, \cdots , \dim M)$ is a general coordinate system on the submanifold. The above equations allow one to determine (up to an $SO(\dim L - \dim M)$ rotation) a part of a Lorentz matrix $\Lambda^{\hat a}{}_{a}(x)$ definable locally everywhere on the submanifold,
\bea
\eta_{\hat a \hat b} \Lambda^{\hat a}{}_{a} \Lambda^{\hat b}{}_{b} = \eta_{a b} ~, &&  \eta^{a b} \Lambda^{\hat a}{}_{a} \Lambda^{\hat b}{}_{b} = \eta^{\hat a \hat b} ~. 
\eea
Here the new index is given by $\hat a = (\al, A)$, $(A = 1, 2, \cdots, \dim L- \dim M)$. The part determined by eq.(\ref{submanifold-eq}) 
is $\Lambda^A{}_{a}$. This, in a sense, defines the basis of normal frames on the submanifold and therefore, given $\Lambda^A{}_{a}$,
\bea
\Lambda'^{A}{}_{a} &=& S^A{}_B \Lambda^B{}_{a} ~, \quad S \in SO(\dim L - \dim M) ~,
\label{Lambda'-Lambda}
\eea
is another good choice\footnote{Note that for a pseudo-Riemannian manifold $S$, in general, is a Lorentz rotation. }. It turns out that all the tubular expansion coefficients of the metric can be written in terms of $f^{a}{}_{, \al} := \del_{\al} f^{a} $, $\Lambda^A{}_{a}$, $ \del_{\al} \Lambda^A{}_{a}$ and certain geometric quantities, namely the connection one form $\omega_{a}$, Riemann curvature $r^{a}{}_{b c d} $ and its higher covariant derivatives - all evaluated on the submanifold in a priori system. We explicitly evaluate all coefficients for vielbein, coefficients of metric up to quartic order and verify correctness of them up to quadratic order.

In addition to this we perform certain higher order verification of the result of \cite{tubular} itself. The result for the transverse components of vielbein is same as that of Riemann normal expansion of \cite{muller}, as expected. However, the new result for the longitudinal components, have not been verified so far beyond quadratic order\footnote{Up to quadratic order, the verification was done in \cite{cut-off}. }. The reason why a higher order verification is important is as follows. As reviewed in Appendix \ref{a:theorem}, the relevant result contains two classes of terms - one independent of spin connection and the other linear in spin connection. The coefficients of these two classes have different closed-form expressions and the first non-trivial term in second class appears at cubic order. In this work we verify this term\footnote{This computation involves manipulating close to $100$ terms containing Christoffel symbols, spin connection and their derivatives. As explained in detail in Appendix \ref{as:cubic}, we adopt a strategy to track various terms to get control over the computation. }. 

In \cite{cut-off}, tubular geometry of loop space $LM$ corresponding to a (preudo)Riemannian manifold $M$ around the submanifold of vanishing loops was computed. This was obtained by taking a suitable large $n$ limit of the tubular geometry around $\Dlt \hookrightarrow M^n$, where $M^n$ is the Cartesian product of $n$ copies of $M$ and $\Dlt \cong M$ is the diagonal submanifold. We show how this constitutes a non-trivial example of our general construction. In spite of this fact the work of \cite{cut-off} was done without the prior knowledge of the present work. We explain how this was possible because of certain specificities of this example.

As another example, we consider the work of Klein and Collas (KC) in \cite{kc-exact} where the complete FNC and metric in FNC were computed exactly in a certain specific situation. We reproduce all these results, some with complete exactness and some up to certain orders, using our method. In particular, our computation verifies a sub-sector of the coefficients for the transverse component to all orders and those in the first class terms for longitudinal component up to quartic order. However, all the spin connection terms in the second class vanish in this specific example, forcing us to perform the aforementioned cubic-order-verification.

The plan for the rest of the paper is as follows. Our entire construction, the indirect method and result for metric expansion (uo to quartic order) are presented in \S \ref{general}. \S \ref{verification} is dedicated to the issue of verification. The two examples, namely $\Dlt \hookrightarrow M^n$ and KC background \cite{kc-exact} are discussed in \S \ref{examples}. We end with an outlook in \S \ref{end}. Many technical details are reported in several appendices.

\section{General tubular expansion problem}
\label{general}

\subsection{The construction} 
\label{s:construction}

\subsubsection{The setup}
\label{ss:setup}

We begin by recalling the following well known description \cite{tubnbh} of an embedded submanifold. We consider an arbitrary submanifold $M$ embedded in a (pseudo)Riemannian ambient manifold $L$. The tangent space to $L$ at 
$Q \in M \hookrightarrow L$ decomposes as follows,
\bea
T_QL = T_QM \oplus N_QM~,
\label{TQL-decom}
\eea
where $N_QM$ is the space of vectors normal to $M$. We then construct the normal bundle $NM$ such that its base is given by $(NM)_0 = M$ and the fiber at $Q\in M$ is given by $N_QM$. A tubular neighborhood around $M\hookrightarrow L$ exists iff it is always possible to find a neighborhood $U \subset L$, with $U \cap M$ non-null, which satisfies the following condition. There exists a convex neighborhood \cite{tubnbh} $\hat U \subset NM$, with $\hat U \cap (NM)_0$ non-null, which is diffeomorphic to $U$ such that $\hat U \cap (NM)_0$ is identically mapped to $U \cap M$. We denote this diffeomorphism by $\Phi$,
\bea
\Phi : \hat U \to U~, \quad \Phi: \hat U \cap (NM)_0 \to U \cap M~, \quad \Phi|_{\hat U \cap (NM)_0} = \hbox{id} ~.
\label{Phi}
\eea
A more explicit description of $\Phi$ can be given as follows. Let us denote an arbitrary element of $\hat U$ by $\hat P= (Q, \hv)$, where $Q \in (NM)_0=M$ and $\hv \in N_QM$. Then the corresponding point $P\in U$ is a unique point on the geodesic starting from $Q \in M\hookrightarrow L$ with initial tangent vector proportional to $\hv \in N_QM$ (i.e. orthogonal to the submanifold). More precisely, $ P = \exp_Q \hv$, where $\exp_Q: T_QL \to L$ is the exponential map of $L$ based at $Q$. The neighborhood $U$ is restricted by the fact that $\exp_Q\hv$, $\forall \hat P = (Q, \hv) \in \hat U $ is a diffeomorphism.

We now translate the above set up in terms of local coordinate systems. The local coordiante system in $U$, i.e. the 
a priori system, is given by (\ref{zbar}). The metric and vielbein components are denoted by $g_{ a b}$ and $e^{(a)}{}_{b}$ respectively,
\bea
\eta_{a b} e^{(a)}{}_{c} e^{(b)}{}_{d} &=& g_{c d}~.
\label{ebar-gbar}
\eea
Finally, the submanifold $(M \cap U) \hookrightarrow U$ is specified by eq.(\ref{submanifold-eq}), with $x^{\alpha}$ denoting a general coordinate system on $M \cap U$. 

Given the above geometric data, the diffeomorphism in (\ref{Phi}) induces a natural Riemannian structure on $NM$. The natural coordinate system in $NM$ is identified as the FNC 
\bea
\hat z^{\hat a} = (x^{\al}, \hat y^A) ~, 
\label{zhat} 
\eea
$x$ being the general coordinate system on the base $(NM)_0 \cap \hat U (= M \cap U)$ and $\hat y$ being coordinates along the fiber. The metric and veilbein components in FNC are denoted as $\hat g_{ab}$ and $\hat e^{(a)}{}_b$ respectively. 

We therefore have the following identifications in terms of coordinates. In our chosen systems, the coordinates of $P \in U$, 
$\hat P \in \hat U $, $Q \in (NM)_0 \cap \hat U$ and $\hv \in N_QM$ are given by $z^{a}$, $\hat z^{\hat a}$, $x^{\al}$ and $\hat y^A$ respectively. The map $\Phi$ is same as the coordinate transformation: $\hat z^{\hat a} \to z^{a} = z^{a}(\hat z)$. More explicitly, this is given by,
\bea
z^{a} &=& f^{a}(x) + \exp_{f(x)}^{a} (\xi) ~, 
\label{z-zhat}
\eea
where,
\bea
\exp_{f(x)}^{a} (\xi) &=& \xi^{a} - \sum_{n\geq 0} {1\over (n+2)!} \underline{ \gamma^{ a}_{ b_1 b_2 \cdots b_{n+2}}} 
\xi^{b_1 \cdots b_{n+2}} ~, \cr
&=& K^{a}{}_B \yh^B - \sum_{n\geq 0} {1\over (n+2)!} \underline{ \gamma^{a}_{b_1 b_2 \cdots b_{n+2}}} K^{b_1}{}_{B_1} 
K^{b_2}{}_{B_2} \cdots K^{b_{n+2}}{}_{B_{n+2}} \yh^{B_1 \cdots B_{n+2}}~. \cr &&
\label{exp}
\eea
The first line in (\ref{exp}) describes the exponential map\footnote{It is well known that the exponential map can be derived for example from the geodesic equation by repeatedly differentiating it. We shall have more detailed comments regarding this in \S \ref{ss:lambda-exp}. } in a priori system, $f^{a}$ and $\xi^{a}$ being the coordinate descriptions for $Q\in M \cap U$ and $\hv \in N_QM$ respectively in the same system and,
\bea
K^{a}{}_{\hat b} &=& \underline{k^{a}{}_{\hat b} }  := \underline{ \lt( \del z^{a} \over \del \hat z^{\hat b} \rt) } ~,
\label{K-matrix}
\eea
is the relevant Jacobian matrix restricted to submanifold. 

Following \cite{cut-off}, we have adopted the following notations. First, the multi-indexed notation for a vector or a coordinate: $\xi^{b_1 \cdots b_n} = \xi^{b_1} \cdots \xi^{b_n}$. We shall use this notation for different variables throughout this article. Second, to reduce clutter we shall usually omit the argument $z$ or $\hat z$ of a quantity, in which case the quantity will be understood to be computed in the tubular neighborhood. An underline will be used to indicate that the quantity is computed on the submanifold, 
i.e at $z = f(x)$ or $\hat z = (x, \hat y =0)$ depending on the coordinate system being used. For example, 
$\underline{\gamma^{ a}_{ b c}} = \gamma^{ a}_{b c} (z = f(x) )$.

The multi-indexed gamma coefficients are symmetric in lower indices and are given by,
\bea
\gamma^{a}_{b_1 b_2 b_3} &=& {1\over 3} \lt( D_{b_1} \gamma^{a}_{b_2 b_3} + \cdots \rt) ~, \cr
\gamma^{a}_{b_1 b_2 b_3 b_4} &=& {1\over 4} \lt( D_{b_1} \gamma^{a}_{b_2 b_3 b_4 } + \cdots \rt) ~, 
\label{multi-gamma}
\eea
and so on. $D_a$ is the covariant derivative in a priori system. In the above equations it acts only on the lower indices \cite{eisenhart} and the ellipses include other terms required for symmetrization.

\subsubsection{The problem}
\label{ss:problem}

We now consider a situation where the local geometry in $U$ is given in the a priori system (\ref{zbar}). i.e. the metric/vielbein as in (\ref{ebar-gbar}) and the equations for submanifold (\ref{submanifold-eq}) are known in this system. The question we would like to pose is how to expand the geometry in small distance from the submanifold. One can of course ordinary Taylor expand 
$g_{a b}$, but in that case the coefficients do not possess nice tensorial properties. The latter, however, is the case when FNC is used. One can therefore proceed in the following way, which we call {\it the direct method}. One first relates the metric in FNC and a priori systems,
\bea
\hat g_{\hat a \hat b} &=& k^{a}{}_{\hat a} k^{b}{}_{\hat b} g_{a b} ~,
\label{ghat-g}
\eea
where both the sides are understood to be expanded in powers of $\hat y$. While the expansion for $k$ is computed from (\ref{z-zhat}), the same for $ g$ is obtained by first ordinary Taylor expanding it as: $ g_{ a  b}(f+  l) = \underline{ g_{ a b}} + \underline{\del_{ c}  g_{ a b}}  l^{ c} + \cdots  $  and then expanding each factor of $ l$ as: $ l^{ a} =  \exp_{f}^{ a} ( \xi)$ and using (\ref{exp}). Clearly, on the RHS of (\ref{ghat-g}) there are three sources of terms that accumulate at each order of $\hat y$. At each order, all these terms added together must reproduce the right tubular expansion coefficient on the LHS expressed in terms of the a priori data. Such a procedure, however, is very cumbersome and goes out of hand in a few orders.  

\subsubsection{The indirect method}
\label{ss:method}

As advocated in \cite{tubular, cut-off}, a lot can be achieved by adopting the {\it indirect method} which uses the results of \cite{tubular}. As summarized in Appendix \ref{a:theorem}, this gives the tensorial characters of all the coefficients for vielbein. In this case, the problem is that the coefficients are known only in FNC. Therefore, our job is to evaluate them in terms of the a priori data. Note that this requires only a small amount of information, as one needs to use the Jacobian matrix evaluated only on the submanifold. This would imply that the only job is to construct the matrix $K$ from (\ref{submanifold-eq}). This however, is not entirely true as, unlike Riemann normal expansion, tubular expansion involves spin connection which gives rise to certain inhomogeneous terms when expressed in terms of a priori data. However, due to the very special way how spin connection appears in the vielbein-expansion, it turns out that we still need a small amount of information. We now proceed to detail our construction below. 

We begin by noting that in the a priori system the internal frames are distributed arbitrarily, while that in the Fermi system are {\it aligned} along the geodesics transverse to the submanifold. We therefore introduce parallel and transverse vielbein components in the a priori system,
\bea
e_{\parallel}^{(\al)}{}_{b} := \lambda^{\al}{}_a e^{(a)}{}_{b} ~, && e_{\perp}^{(A)}{}_{b} := \lambda^A{}_{a} e^{(a)}{}_{b} ~,
\label{e-lambda-def}
\eea
where $\lambda$ is a suitable Lorentz transformation matrix defined locally everywhere,
\bea
\eta_{\hat a \hat b} \lambda^{\hat a}{}_{a} \lambda^{\hat b}{}_{b} = \eta_{a b} ~,  && \eta^{a b} \lambda^{\hat a}{}_{a} \lambda^{\hat b}{}_{b} = \eta^{\hat a \hat b} ~.
\label{lambda-square}
\eea
We shall also use the notation $e_{\lambda}^{(\hat a)}{}_{b}$ to collectively denote 
$\{ e_{\parallel}^{(\al)}{}_{b}, e_{\perp}^{(A)}{}_{b} \}$. 

The fact that $e_{\lambda}^{(\hat a)}{}_{b}$ are aligned frames means the following. Let us consider the transverse geodesic with initial tangent vector proportional to $\hv \in N_QM$ as explained below eq.(\ref{Phi}). We introduce a parameter $t$ along the geodesic such that at the starting point $Q \in M \cap U$, $t=0$. A priori coordinates for the point at $t$ is given by, $z^a(t) = f^a(x) + \exp^a_{f(x)}(\xi(t)) $ where $\xi^a(t)$ are the components of $\hv (t) \in N_QM$ in the same system with initial value given by, $ \xi^{ a}(0) =  \xi^{ a}$. Then the components in internal frame are given by\footnote{With an abuse of language, by $ e^{( a)}{}_{ b} (t)$ we actually mean $ e^{( a)}{}_{ b} (z(t))$. },
\bea
 \zeta^{( a)} (t) &=&  e^{( a)}{}_{ b} (t)  \xi^{ b} (t) ~.
\label{zeta-xi}
\eea
Since the vielbein in the a priori system is arbitrary, the internal frame components of the tangent vactor changes along the geodesic. However, the components along the aligned frames, namely,
\bea
\hat \zeta^{(A)} :=  e_{\lambda}^{(A)}{}_{ b}  \xi^{ b}(t) = \lambda^A{}_{ a} (t)  \zeta^{( a)}(t)  ~,
\label{zeta-hat-zeta}
\eea
remain constant along the geodesic.

We can now specify the relation between the vielbein components in Fermi and a priori systems. It is given by,
\bea
\hat e^{(\hat a)}{}_{\hat b} &=& k^{ b}{}_{\hat b}  e_{\lambda}^{(\hat a)}{}_{ b} = \lambda^{\hat a}{}_{ a} k^{ b}{}_{\hat b}  
e^{( a)}{}_{ b} ~.
\label{ehat-e}
\eea
This is the key equation that guides us to implement the indirect method (see below), as well as to verify consistency of the whole construction. Once we know how to expand the $\lambda$-matrix in powers of $\hat y$, the expansion of the entire expression on the RHS can be evaluated following the same procedure as described below eq.(\ref{ghat-g}). Consistency would then require that at each order, this computation must reproduce the same tubular expansion coefficients as described in Appendix \ref{a:theorem}, expressed in terms of a priori data. According to the indirect method, for the purpose of evaluating the final result this entire computation can be avoided simply by directly writing these coefficients in terms of a priori data using their transformation laws. The small amount of {\it local data} that is required for this purpose are the $K$-matrix in (\ref{K-matrix}) and 
$\Lambda^{\hat a}{}_{ b} := \underline{ \lambda^{\hat a}{}_{ b} }$. While all the tensors are transformed by the $K$-matrix, the transformation of spin connection, restricted to the submanifold takes the following form,
\bea
\underline{ \hat \omega_{\hat a}{}^{(\hat b)}{}_{(\hat c)}  } 
&=& \lt\{ \begin{array}{ll}
K^{ a}{}_{ \al} \Lambda^{\hat b}{}_{ b} \Lambda_{\hat c}{}^{ c} \underline{  \omega_{ a}{}^{( b)}{}_{( c)} } 
+ \Lambda^{\hat b d} \del_{\al} \Lambda_{\hat c d}  & \quad \hbox{for } \hat a = \al \cr & \cr
0  & \quad \hbox{for } \hat a = A ~,
\end{array} \rt. 
\label{omg-hat-sub}
\eea
where we raise or lower the indices of $\Lambda$ (or $\lambda$) by $\eta$.

We now proceed to construct the aforementioned local data in terms of the a priori data. The map in (\ref{z-zhat}) implies,
\bea
K^{ a}{}_{\hat b} &=& (f^{ a}{}_{,\beta}, K^{ a}{}_B ) ~, 
\label{K-a-b-hat}
\eea
where $K^{ a}{}_B$ is yet to be determined. This can be fixed once the following requirement (part of the definition of FNC \cite{tubular}) is imposed, 
\bea
\underline{\hat e^{(\hat a)}{}_{\hat b} } &=& diag(E^{(\al)}{}_{\beta}, \eta^A{}_B)~,
\label{ehat-sub}
\eea
where $E^{(\al)}{}_{\beta}(x)$ is the induced vielbein on submanifold\footnote{ Therefore,
\bea
G_{\al \beta} = f^{ a}{}_{,\alpha} f^{ b}{}_{, \beta} \underline{ g_{ a  b}} = E^{(\gamma)}{}_{\al} E^{(\dlt)}{}_{\beta} \eta_{\gamma \dlt}~, 
\eea
is the induced metric.}. Below we shall argue that the solution for $K$ is given by,
\bea
K^{ a}{}_B = \underline{ e_{\lambda (B)}{}^{ a}}~.
\label{K-a-B}
\eea 

We begin by computing the matrix elements $\Lambda^A{}_b$. To this end we define the following $(\dim L \times \dim M)$ matrix,
\bea
\Omg_{ a \alpha} &=& \underline{ g_{ a  b} } f^{ b}{}_{,\alpha}  ~, \quad \hbox{rank $\Omg$} = \dim M~.
\eea
Given any vector $\hv \in N_QM$, its components in a priori system must satisfy,
\bea
 \xi^{ a} \Omega_{ a \al} = 0~,
\label{xi-eq}
\eea
which admit $(\dim L - \dim M)$ number of independent solutions. Any such solution can be expanded linearly in terms of the vielbein components. An orthonormal set $\{ \xi_{(A)}(x)\}$ is given by,
\bea
 \xi^{ a}_{(A)} &=& \underline{ e_{\lambda (A)}{}^{ a}} = \Lambda_A{}^{ b} \underline{ e_{( b)}{}^{ a}} ~.
\label{xi-sol}
\eea
Once all the independent solutions of (\ref{xi-eq}) are found, the matrix elements $\Lambda^A{}_b$ can be computed using the above equation. 

Given the above result, we now go back to the solution (\ref{K-a-B}) and check using the transformation law (\ref{ehat-e}) that the expected result in (\ref{ehat-sub}) is reproduced. 
\bea
\underline{\hat e^{(\al)}{}_{\beta}} = f^{ c}{}_{,\beta} \underline{ e_{\lambda }^{(\al)}{}_{ c}} =: E^{(\al)}{}_{\beta} ~, && \quad
\underline{\hat e^{(A)}{}_{\beta}}  = f^{ c}{}_{, \beta} \underline{ e_{\lambda }^{(A)}{}_{ c}} = \eta^{AB}  \xi^{ c}_{(B)} \Omg_{ c \beta} = 0 ~, \\
\label{ehat-a-beta-sub} 
\underline{\hat e^{(\al)}{}_B} = \underline{ e_{\lambda (B)}{}^{ c}  e_{\lambda }^{(\al)}{}_{ c}} = 0~, && \quad 
\underline{\hat e^{(A)}{}_B} = \underline{ e_{\lambda (B)}{}^{ c}  e_{\lambda }^{(A)}{}_{ c}} = \dlt^A{}_B~,
\label{ehat-a-B-sub}
\eea
which, by eq.(\ref{ehat-sub}), are the expected results. One may also check that $E^{(\alpha)}{}_{\beta}$, as defined above, does indeed qualify for the induced vielbein on $M$,
\bea
\eta_{\alpha \beta} E^{(\alpha)}{}_{\gamma} E^{(\beta)}{}_{\dlt} &=& \eta_{\alpha \beta} (f^{ a}{}_{,\gamma} \underline{ e_{\lambda}^{(\alpha)}{}_{ a}} ) (f^{ b}{}_{,\dlt} \underline{ e_{\lambda }^{(\beta)}{}_{ b}} ) ~, \cr
&=& \eta_{\alpha \beta} (f^{ a}{}_{,\gamma} \underline{ e_{\lambda }^{(\alpha)}{}_{ a}} ) 
(f^{ b}{}_{,\dlt} \underline{ e_{\lambda }^{(\beta)}{}_{ b}} ) 
+ \eta_{AB} (f^{ a}{}_{,\gamma} \underline{ e_{\lambda }^{(A)}{}_{ a}} ) 
(f^{ b}{}_{,\dlt} \underline{ e_{\lambda }^{(B)}{}_{ b}} ) ~, \cr
&& \lt[ \hbox{ The second term being zero } \rt] ~, \cr
&=& f^{ a}{}_{,\gamma}  f^{ b}{}_{,\dlt} \underline{ g_{ a  b}} = G_{\gamma \dlt}(x)~.
\eea

\subsubsection{Non-coordinate frame analogue of (derivative of) exponential map}
\label{ss:lambda-exp} 

Verifying the consistency condition, as explained below eq.(\ref{ehat-e}), is an important step of our construction. While the verification will be performed in \S \ref{verification}, here we shall derive expansion for the $\lambda$-matrix. 

To this end we first go back to considering the transverse geodesic parametrized by $t$ as discussed above eq.(\ref{zeta-xi}). The tangent vector satisfies the following relation,
\bea
 \xi^{ a}(t) &=& {d  z^{ a} (t) \over dt}~.
\eea
This can be written in terms of the initial tangent vector $ \xi^{ a}$ as follows,
\bea
 \xi^{ a}(t) &=& \phi^{ a}{}_{ b} (t)  \xi^{ b} ~, \cr
\phi^{ a}{}_{ b} (t) &=& \dlt^{ a}{}_{ b} - \underline{ \gamma^{ a}_{ c_1  b }}  \xi^{ c_1} t 
- {1\over 2} \underline{ \gamma^{ a}_{ c_1  c_2  b }}  \xi^{ c_1  c_2} t^2 
- {1\over 3!} \underline{ \gamma^{ a}_{ c_1  c_2  c_3  b }}  \xi^{ c_1  c_2  c_3 } t^3 - \cdots ~,
\label{derv-exp}
\eea
where $\phi^a{}_b(t)$ is obtained by differentiating the exponential map in (\ref{exp}). We now relook at eq.(\ref{zeta-hat-zeta})
by inverting it,
\bea
\zeta^{(a)} (t) &=& \theta^a{}_{\hat b}(t) \hat \zeta^{(\hat b)} ~, \quad \quad [ \hat \zeta^{(\beta)} = 0 \hbox{ and } \theta^a{}_{\hat b} = \lambda_{\hat b}{}^a ] 
\label{zeta-zeta-hat}
\eea
Comparing this equation with the first equation in (\ref{derv-exp}) one concludes that $\theta(t)$ is the non-coordinate frame analogue of 
$\phi(t)$ and one would like to find it's expanded form analogous to the second equation in (\ref{derv-exp}).  

At this point it is useful to recall that the second equation in (\ref{derv-exp}) can alternatively be derived by repeatedly differentiating the geodesic equation \cite{eisenhart}\footnote{In fact the standard method is to first derive (\ref{derv-exp}) from geodesic equation and then to find the exponential map in (\ref{exp}) by integrating (\ref{derv-exp}). \label{f:exp-der}},
\bea
 D_{ \xi(t)}  \xi(t) &=& 0~, 
\label{geodesic-coord1}
\eea
which, in the coordinate frame, takes the following form,
\bea
{d  \xi^a(t) \over dt} +  \gamma^a_{bd}  \xi^b(t)  \xi^d(t) &=& 0~.  
\label{geodesic-coord2}
\eea
To find the non-ccordinate frame analogue of this procedure, we re-write the LHS of (\ref{geodesic-coord1}) in the following manner,
\bea
 D_{ \xi(t)}  \xi(t) &=&  \xi^{ a} (t)  D^{t}_{ a}  \zeta^{( b)} (t)  {\cal E}_{( b)} (t) ~,
\label{D-xi-internal}
\eea
where ${\cal E}_{( a)} (t) =  e_{( a)}{}^{ b} (t) \del_{ b} $ are the non-coordinate frames. $D^{t}_a$ is the total covariant derivative (in a priori system) which annihilates the vielbein,
\bea
D^{t}_a e^{(b)}{}_c &=& D_a e^{(b)}{}_c + \omg_a{}^{(b)}{}_c ~, 
\eea
where $\omg_a{}^{(b)}{}_{(c)}$ are the spin connection coefficients and we have converted a Lorentz index to coordinate index using vielbein - a convention that will be used throughout this article (except for certain multi-indexed $\omg$-symbols that will appear below). Rewriting eq.(\ref{geodesic-coord1}) using (\ref{D-xi-internal}) leads to the following equation, 
\bea
{d  \zeta^{( b)} \over dt} +  \xi^{ a}(t)  \omega_{ a}{}^{( b)}{}_{( c)}  \zeta^{( c)}(t) &=& 0~.
\eea
Now substituting (\ref{zeta-zeta-hat}) into the above equation and demanding that $\hat \zeta^{(\hat b)}$ be independent of $t$, one arrives at the following equation,
\bea
{d \over dt} \theta^{ a}{}_{\hat b} (t) +  \xi^{ c} (t)  \omega_{ c}{}^{( a)}{}_{( e)} \theta^{ e}{}_{\hat b} (t) &=& 0~.
\eea
To solve this equation we follow through the same procedure as used to find (\ref{derv-exp}). By repeatedly differentiating the above equation and using (\ref{geodesic-coord2}) one gets the following result,
\bea
{d^n \over dt^n} \theta^{ a}{}_{\hat b} (t) &=& -  \omega_{ c_1 \cdots  c_n}{}^{( a)}{}_{( d)} \theta^{ d}{}_{\hat b} (t)  \xi^{ c_1 \cdots  c_n}(t)  ~, 
\eea
where the multi-indexed spin connection coefficient (analogue of the $ \gamma$-coefficients in (\ref{exp})) is symmetric in its coordinate indices and is define by the following recurrence relation,
\bea
\omg_{ c_1  \cdots c_n }{}^{( a)}{}_{( b)} &=& {1\over n} \lt[ D^{t}_{c_1} \omg_{c_2 \cdots c_n}{}^{(a)}{}_{(b)} 
- \omg_{c_1}{}^{(a)}{}_{(e)} \omg_{c_2 \cdots c_n}{}^{(e)}{}_{(b)} + \cdots \rt]~, \quad \forall n > 1 ~.
\label{multi-omega}
\eea
The ellipses contain other terms required for symmetrisation among the coordinate indices. Note that according to our definition, unlike the $\gamma$-coefficients, the total covariant derivative in the above equation applies on all the indices.

This leads to the following solution for 
$\lambda$-matrix, 
\bea
\lambda^{\hat a}{}_{ b} &=&  \Lambda^{\hat a}{}_{ b'} \lt[ \dlt^{ b'}{}_{ b} - \underline{ \omega_{ c_1( b)}{}^{( b')}}  \xi^{ c_1} 
- {1\over 2} \underline{  \omega_{ c_1  c_2 ( b)}{}^{( b')}  }  \xi^{ c_1  c_2}  
- {1\over 3!} \underline{ \omega_{ c_1  c_2  c_3 ( b)}{}^{( b')} }  \xi^{ c_1  c_2  c_3} - \cdots \rt] ~, \cr && 
\label{lambda-exp}
\eea
where we have chosen to write the result in terms of $  \xi^{ a} = K^{ a}{}_B \hat y^B$ instead of $\hat y$ for notational simplicity. We have also checked explicitly up to cubic order that the above solution satisfies the identities in (\ref{lambda-square}).

We end this subsection with certain remarks regarding alignment of internal frames as defined near eq.(\ref{zeta-zeta-hat}) in light of the above result. Equation (\ref{lambda-exp}) implies that a given internal frame is aligned iff the following conditions are satisfied,
\bea
\underline{\omg_{c_1 \cdots c_n}{}^{(a)}{}_{(b)} } \xi^{c_1 \cdots c_n} &=& 0~, \quad \forall n \geq 1~,
\label{alignment-cond}
\eea
which is a coordinate independent statement, as it should be. Notice also that conditions in (\ref{FNC-cond}) for FNC not only choose a special coordinate system, but also make the internal frame aligned. This is because the second equation in (\ref{FNC-cond}) guarantee, as can be argued using eq.(\ref{multi-der-omg}), that the above conditions are satisfied. It is therefore obvious that any system that is obtained by giving a coordinate transformation on FNC without altering the internal frame, remains aligned. This point will be crucial in our discussion in \S \ref{ss:cut-off}.

\subsection{Result for metric expansion up to quartic order}
\label{s:metric}

Below we explicitly write down the expansion for metric in terms of the a priori data up to quartic order.  
\bea
\hat g_{\al \beta} 
&=& G_{\al \beta} + \lt[ f^{ a}{}_{, \al} f^{ b}{}_{, \beta} \Lambda_C{}^{ c} \underline{ \omega_{ a  b ( c)}  } + f^{ b}{}_{, \beta} \del_{\al}  \Lambda_C{}^{ c } \underline{  e_{( c )  b} }  + \al \leftrightarrow \beta \rt] \hat y^C \cr
&& + \lt[ f^{ a}{}_{, \al} f^{ b}{}_{, \beta } \Lambda^{ c_1}_{C_1}{}^{ c_2}_{ C_2} 
\lt( \underline{ r_{ a ( c_1  c_2)  b} } + \underline{  \omega_{ a}{}^{( e)}{}_{( c_1) }  \omega_{ b}{}_{( e  c_2 ) } }  \rt)  \rt. \cr
&& \lt.
- \lt(f^{ a}{}_{, \al} \Lambda^{ c_1}_{C_1}{}_{\beta}{}^{ c_2}_{C_2}  \underline{  \omega_{ a (  c_1  c_2) } }  + \al \leftrightarrow \beta \rt) 
+ \Lambda_{\al}{}^{ c_1}_{C_1}{}_{\beta}{}^{  c_2}_{C_2 } \eta_{ c_1  c_2} \rt] \hat y^{C_1 C_2} \cr
&& + \lt[ f^{ a}{}_{, \al} f^{ b}{}_{, \beta} \Lambda^{ c_1}_{C_1}{}^{ c_2}_{C_2}{}^{ c_3}_{C_3 }
\lt( {1\over 3} \underline{  D^{t}_{( c_1) }  r_{ a ( c_2  c_3)  b } }  
+ {2\over 3} \underline{  \omega_{ a}{}^{( e)}{}_{( c_1 )}  r_{( e c_2  c_3)  b} } 
+ {2\over 3} \underline{  \omega_{ b}{}^{( e)}{}_{( c_1 )}  r_{( e c_2  c_3)  a} }  
\rt)  \rt. \cr
&& \lt. + {2\over 3} \lt( f^{ a }{}_{, \al } \Lambda^{ c_1}_{C_1}{}^{ c_2}_{C_2}{}_{\beta}{}^{ c_3}_{C_3 }  
\underline{  r_{ a ( c_1  c_2  c_3 )  } } + \al \leftrightarrow \beta \rt)  \rt]  \hat y^{C_1 C_2 C_3}  \cr
&& + \lt[  f^{ a}{}_{, \al} f^{ b}{}_{, \beta } \Lambda^{ c_1}_{C_1}{}^{ c_2}_{C_2}{}^{ c_3}_{C_3}{}^{ c_4}_{C_4} 
\lt\{ {1\over 12} \underline{  D^{t}_{( c_1)}  D^{t}_{( c_2)}  r_{  a ( c_3  c_4)  b }  } 
+ {1\over 3} \underline{  r_{ a ( c_1  c_2)  e}  r^{ e}{}_{ ( c_3  c_4)  b } } 
\rt. \rt. \cr
&& \lt.
+ {1\over 4} \lt( \underline{  \omega_{ a}{}^{( e)}{}_{( c_1 )}  D^{t}_{( c_2)}  r_{( e  c_3  c_4)  b} } +  a \leftrightarrow  b \rt)  
+ {1\over 3} \underline{ \omega_{ a}{}^{( e)}{}_{( c_1)}  \omega_{ b}{}^{( f)}{}_{( c_2)}  r_{( e  c_3  c_4  f)} }  \rt\} \cr
&&
+ {1\over 4} \lt( f^{ a}{}_{, \al } \Lambda^{ c_1}_{C_1}{}^{ c_2}_{C_2}{}^{ c_3}_{C_3}{}_{\beta}{}^{ c_4}_{C_4} \underline{  D^{t}_{( c_1)}  r_{  a ( c_2  c_3  c_4 )} }  + \al \leftrightarrow \beta \rt) 
\cr
&& 
+ {1\over 3} \lt( f^{ a}{}_{, \al} \Lambda^{ c_1}_{C_1}{}^{ c_2}_{C_2}{}^{ c_3}_{C_3}{}_{\beta}{}^{ c_4}_{C_4}
\underline{ \omega_{ a}{}^{( e)}{}_{( c_1)}  r_{( e  c_2  c_3  c_4 )}   } 
+ \al \leftrightarrow \beta  \rt)  \cr
&& \lt.
+ {1\over 3} \Lambda_{\al}{}^{ c_1}_{C_1}{}^{ c_2}_{C_2}{}^{ c_3}_{C_3}{}_{\beta}{}^{ c_4}_{C_4} \underline{  r_{( c_1  c_2  c_3  c_4 )} }   \rt] \hat y^{C_1 \cdots C_4}  + O(\hat y^5)  ~, 
\label{lhs-g-al-beta} \\
\hat g_{\al B} 
&=& \lt( f^{ a}{}_{, \al} \Lambda^{ b}_B{}^{ c}_C \underline{ \omega_{ a ( b  c) }} 
+ \Lambda^{ b}_B{}_{\al}{}^{ c}_C\eta_{ b  c} \rt) \hat y^C 
+ {2\over 3} f^{ a}{}_{, \al}  \Lambda^{ c_1}_{C_1}{}^{ c_2}_{C_2}{}^{ b}_B
\underline{  r_{ a ( c_1  c_2  b) } } \hat y^{C_1 C_2}  \cr
&& + \lt[f^{ a}{}_{, \al} \Lambda^{ c_1}_{C_1}{}^{ c_2}_{C_2}{}^{ c_3}_{C_3}{}^{ b}_B 
\lt(  {1\over 4}  \underline{ D^{t}_{( c_1)}  r_{ a (  c_2  c_3  b) }} 
+ {1\over 3} \underline{  \omega_{ a}{}^{( e)}{}_{( c_1)}  r_{( e  c_2  c_3  b) } }  \rt)  \rt. \cr
&& \lt.
+ {1\over 3} \Lambda_{\al}{}^{ c_1}_{C_1}{}^{ c_2}_{C_2}{}^{ c_3}_{C_3}{}^{ b}_B
\underline{  r_{( c_1  c_2  c_3   b)} }  \rt] \hat y^{C_1 C_2 C_3}  \cr
&& + \lt[ f^{ a}{}_{, \al} 
\Lambda^{ c_1}_{C_1}{}^{ c_2}_{C_2}{}^{ c_3}_{C_3}{}^{ c_4}_{C_4}{}^{ b}_B
\lt( {1\over 15} \underline{ D^{t}_{( c_1)}  D^{t}_{( c_2)}   r_{ a ( c_3  c_4  b) }} 
+ {2\over 15} \underline{ r_{ a ( c_1  c_2) e}  r^{e}{}_{( c_3  c_4  b) }} \rt. \rt. \cr
&& \lt. \lt.
+ {1\over 6}  \underline{  \omega_{ a}{}^{( e)}{}_{( c_1)}  D^{t}_{( c_2)}  r_{( e  c_3  c_4  b) } }  \rt)  
+ {1\over 6 }  \Lambda^{ c_1}_{C_1}{}_{\al}{}_{ C_2}^{ c_2}{}^{ c_3}_{C_3}{}^{ c_4}_{C_4}{}^{ b}_B
\underline{  D^{t}_{( c_1)}  r_{( c_2  c_3  c_4  b)} } \rt] \hat y^{C_1 \cdots C_4}  + O(\hat y^4) ~, \cr
&&
\label{lhs-g-al-B} \\
\hat g_{AB} 
&=& \eta_{AB} + {1\over 3 } \Lambda^{ a}_A{}^{ c_1 }_{C_1}{}^{ c_2}_{C_2}{}^{ b }_B \underline{ r_{ ( a  c_1  c_2  b) } } \hat y^{C_1 C_2}
+ {1\over 6 } \Lambda^{ a}_A{}^{ c_1 }_{C_1}{}^{ c_2}_{C_2}{}^{ c_3}_{C_3}{}^{ b }_B \underline{  D^{t}_{( c_1)}  r_{ ( a  c_2  c_3  b) }  } \hat y^{C_1 C_2 C_3} \cr
&& + \Lambda^{ a}_A{}^{ c_1 }_{C_1}{}^{ c_2}_{C_2}{}^{ c_3}_{C_3}{}^{ c_4}_{C_4}{}^{ b }_B 
\lt[ {1\over 20} \underline{ D^{t}_{( c_1)}  D^{t}_{( c_2) }  r_{( a  c_3  c_4  b ) }} + {2\over 45} \underline{ r_{( a  c_1  c_2) e}  r^{e}{}_{( c_3  c_4  b) }} \rt] \hat y^{C_1 \cdots C_4} 
+ O(\hat y^4) ~, \cr
&&
\label{lhs-g-A-B}  
\eea
where we have used notations as defined in (\ref{notations}). Furthermore, a lower parallel index $(\al)$ in a $\Lambda$-symbol indicates a parallel derivative acting on one of its 
$\Lambda$-factors depending on its position. For example,
\bea
\Lambda^{\cdots  c_1}_{\cdots C_1 }{}_{\al}{}^{ c_2 }_{C_2}{}^{ c_3 \cdots}_{C_3 \cdots} &=& \cdots \Lambda_{C_1}{}^{ c_1} \del_{\al} \Lambda_{C_2}{}^{ c_2} \Lambda_{C_3}{}^{ c_3} \cdots ~.
\eea
The details of derivation of the above results are given in Appendix \ref{a:comp}. Notice that, as mentioned below eq.(\ref{Lambda'-Lambda}), the only local data needed to write the metric expansion are $f^a{}_{, \al}$ and the matrix elements $\Lambda^A{}_b$, both to be obtained from the submanifold equation (\ref{submanifold-eq}). 

\section{Verification}
\label{verification}

Here we address the problem of verification. The goal is twofold, which we discuss below separately.

\subsection{Consistency of general construction }  
\label{s:consistency}

As mentioned earlier, though the rules of computing tubular expansion of vielbein in terms of a priori data using indirect method are simple, the overall understanding and consistency of the general construction rely on the fact that eq.(\ref{ehat-e}) hold true to all orders. Our goal here is to verify this explicitly up to quadratic order. 

The method of verification is as follows. The expansion on the LHS of eq.(\ref{ehat-e}) is known from the results of \cite{tubular}. We first rewrite these results in terms of the a priori data using our indirect method. The RHS, on the other hand, are ordinary Taylor expanded as explained below equations (\ref{ghat-g}) and (\ref{ehat-e}). Equating these results at each order gives rise to certain identities that need to be satisfied. The details of this procedure have been discussed in Appendix \ref{a:verification}. The non-trivial identities that one arrives at up to quadratic order are given by the first equation in (\ref{omg-hat-sub}) and,
\bea
{1\over 3}  r^{( a)}{}_{ c_1  c_2  b}  \xi^{c_1 c_2} &=& \lt[  e^{( a)}{}_{ b,  c_1  c_2 } 
-  \gamma^{ d}_{ c_1  c_2}  e^{( a)}{}_{  b,  d } -  \gamma^{ d}_{  b  c_1  c_2 }  e^{( a)}{}_{ d} + 2  \omega_{ c_1}{}^{( a)}{}_{( d)}  e^{( d)}{}_{ b,  c_2 }  
\rt. \cr
&& \lt. 
-  \omega_{  c_1  c_2}{}_{( d)}{}^{( a )}  e^{( d)}{}_{ b}  \rt]  \xi^{c_1 c_2} ~, \cr 
 r^{( a)}{}_{ c_1  c_2  b }  \xi^{ c_1  c_2} &=& 
\lt[ e^{( a)}{}_{ b,  c_1  c_2 }  - \gamma^{ d }_{ c_1  c_2}  e^{( a)}{}_{ b,  d } - 2 \gamma^d_{b c_2} e^{(a)}{}_{d,c_1}  
- \gamma^{ d }_{ c_1  c_2,  b }  e^{( a)}{}_{ d}   \rt. \cr
&& \lt. 
+ 2 e^{(a)}{}_d \gamma^d_{c_1 e} \gamma^e_{c_2 b} - 2 \omg_{c_1}{}^{(a)}{}_{(d)} \omg_{c_2}{}^{(d)}{}_b 
- \omega_{ c_1  c_2}{}_{( d )}{}^{( a )}  e^{( d )}{}_{ b}  \rt]  \xi^{ c_1  c_2}  ~. 
\eea
By using techniques of tensor calculus similar to those used in \cite{cut-off} we have explicitly checked the above identities to be true.

\subsection{Result of \cite{tubular} at cubic order } 
\label{s:cubic}

The closed form expression of the tubular expansion coefficients for vielbein are given in Appendix \ref{a:theorem}. This was obtained in \cite{tubular} from certain integral theorem derived in the same work. Although the latter was shown to be consistent with the metric-intergral-theorem of \cite{FS}, so far there has not been any verification of the closed form expressions beyond quadratic order \cite{cut-off}. As mentioned earlier, this tests the first class terms that are independent of spin connection. The first non-trivial spin connection term appears at cubic order. Here our goal is to verify this term. This gives the first non-trivial test for the closed form expression for the second class terms. 

This computation is very long and tedious as the number of relevant terms on the RHS of (\ref{ehat-e}) is nearly $100$. However, we use certain tricks in order to achieve control over the computation. The details of this analysis are presented in Appendix \ref{as:cubic}.

\section{Examples }
\label{examples}

Here we discuss two examples. In \S \ref{s:diagonal} we consider the work of \cite{cut-off}, where tubular geometry around 
$\Dlt \hookrightarrow L=M^n$ was computed using a similar indirect method. Here $M^n$ is the Cartesian product of $n$ copies of $M$ and $\Dlt \cong M$ is the diagonal submanifold. It is natural to ask how this work can be understood as a special case of our present construction. As we shall see, this is indeed a non-trivial special case. However, because of certain specificities of the problem, the method of \cite{cut-off} worked without the knowledge of the general construction. We explain all the subtleties involved. In \S \ref{s:KC} we consider the work of \cite{kc-exact}, where a class of backgrounds was studied for which exact FNC was constructed and exact expression for metric was derived. We shall consider the same class of backgrounds in arbitrary $d$-dimensions and reproduce these results, some exactly to all orders and some up to few non-trivial orders, using our construction. 

In both the examples, however, the $\Lambda$-mtarix is constant over submanifold. This implies that all the $\del_{\al} \Lambda$ dependent inhomogeneous terms in tubular expansion are absent. This makes the verification of our general construction up to quadratic order as done in \S \ref{s:consistency} more so important.

\subsection{$\Dlt \hookrightarrow \cM^n$}
\label{s:diagonal}

In order to closely relate to the work of \cite{cut-off}, we adopt the following change of notations (only for this subsection). The a priori and Fermi systems will be denoted as $\bar z^{\bar a}$ and $\hat z^a = (x^{\al}, \hat y^A)$ respectively. This implies the following change of notation for the indices from the rest of the article: $a \to \bar a$, $\hat a \to a$.

The plan of this subsection is as follows: In \S \ref{ss:M-data} we recall how the geometric quantities in the a priori system are expressed in terms of the geometric quantities of $M$, hereafter to be called $M$-data. These results will be used in the rest of the subsection. Then in \S \ref{ss:specialization} we discuss how $\Dlt \hookrightarrow M^n$ should be interpreted as a special case of the general construction and show that because of curvature of $M$, this is indeed a non-trivial special case. Finally, in \S \ref{ss:cut-off} we elaborate on the coordinate transformation performed in \cite{cut-off} and explain how the method works without the knowledge of general construction.  

\subsubsection{A priori system and $M$-data}
\label{ss:M-data}

The a priori system is given by the direct product coordinates DPC \cite{cut-off},
\bea
\bar z^{\bar a} &=& (x_1^{\al_1}, x_2^{\al_2}, \cdots , x_n^{\al_n}) ~, \quad  \bar a = (\al_1, \al_2, \cdots , \al_n) ~,
\label{zbar-M^n}
\eea
where $x_p^{\al_p}$ $(p=1, \cdots , n)$ is a local coordinate system in the $p$-th copy of $M$, chosen in such a way that the metric in $M^n$ is given by,
\bea
\bar g_{ \bar a  \bar b}(\bar z) d \bar z^{ \bar a} d \bar z^{ \bar b} &=& {1\over n} \sum_{p=1}^n G_{\al_p \beta_p}(x_p) dx^{\al_p} dx^{\beta_p} ~,
\eea
where $G_{\al \beta}$ is the metric in $M$. As explained in \cite{cut-off}, any geometric quantity, say $T$ in $M$, constructed out of vielbein and its derivatives, gives rise to a corresponding geometric quantity $ \bar t$ in $M^n$ in the 
a priori system which is expressed in terms of $T$ in the following way. Let the index structure of $T$ be as follows: $T^{\check \al \check \beta \cdots }_{\check \gamma \check \dlt \cdots }$, where a $\check{}$ on an index indicates that it can either be a coordinate or non-coordinate index. Then one first defines the following geometric quantity in $M^n$,
\bea
 \bar t'{}^{\check{ \bar a} \check{ \bar b} \cdots}_{\check{\bar c} \check{\bar d} \cdots } ( \bar z) &=& \lt\{ 
\begin{array}{ll}
t'_p{}^{\check{\al}_p \check{\beta}_p \cdots}_{\check{\gamma}_p \check{\dlt}_p \cdots }  = T^{\check{\al}_p \check{\beta}_p \cdots}_{\check{\gamma}_p \check{\dlt}_p \cdots } (x_p) ~, 
& \hbox{for } \check{ a} = \check{\al}_p~, \check{ c} = \check{\gamma}_p ~, \cdots  \cr
0 & \hbox{otherwise }~.
\end{array} \rt. 
\label{tbar-prime-T}
\eea  
Note that it is block diagonal in the multi-dimensional sense and that is because we are useing DPC. Finally, $\bar t$, which has the same index-structure as $\bar t'$, is given by,
\bea
 \bar t &=& n^{w\over 2}  \bar t'~,
\eea
where $w$ is the Weyl-weight of $T$. The above definition also works backwords, i.e. given any geometric quantity $ \bar t$, constructed out of vielbein and its derivatives, there exists the corresponding quantity $T$ in $M$, in terms of which $ \bar t$ can be expressed following the above rule. This completes the description of the local geometry of $M^n$ in a priori system in terms of $M$-data. 

\subsubsection{ $\Dlt \hookrightarrow M^n$ as a special case}
\label{ss:specialization}

We now proceed to apply our general construction in this case. The equations that determine the submanifold are given by,
\bea
 x_p^{\al_p} = f^{\al_p }(x) = \dlt^{\al_p}{}_{\al} x^{\al} ~.
\eea
Notice that the general system $x^{\al}$ on the submanifold $\Dlt \cong M$ is chosen in such a way that the induced metric 
$f^{\bar a}{}_{,\al} f^{\bar b}{}_{, \beta} \underline{\bar  g_{\bar a \bar b} }$ is still given by $G_{\al \beta}(x)$. The analogue of eq.(\ref{xi-eq}) to be satisfied by the transverse vector,
\bea
\bar \xi^{\bar a} &=& (\xi_1^{\al_1}, \xi_2^{\al_2}, \cdots ) ~, \quad \xi_p\in T_{x_p=x}M ~,
\label{xibar-M^n}
\eea
can be written as,
\bea
\sum_p \dlt^{\al}{}_{\al_p} \xi_p^{\al_p}(x) &=& 0~.
\label{xi-bar-eq-M^n}
\eea
Solutions to the above equation are given by (\ref{xi-sol}) with the following interpretation. The aligned index is given by:
$a = (\al, A)$ where $A$ is given by a pair: $A = (\hat \al, \ha)$ with $\ha = 1, 2, \cdots (n-1)$ and $\hat \al$ a tangent space index of $M$. Then the $\Lambda$-matrix reads,
\bea
\Lambda^{\al}{}_{\bar b} = O_{0 p} \eta^{\al}{}_{\beta_p} ~, && \Lambda^A{}_{\bar b} =  O_{\ha p} \eta^{\hat \al}{}_{\beta_p}~, \quad 
\hbox{ when } \bar b = \beta_p ~, 
\label{Lambda-M^n}
\eea
where $O \in O(n)$ such that,
\bea
O_{0p} &=& {1\over \sqrt{n}} ~.
\eea
As expected, this definition of $\Lambda$ satisfies eq.(\ref{lambda-square}), with the following interpretation of indices,
\bea
\eta_{ a  b} = \eta_{\al_p \beta_q} \dlt_{p,q} ~, && \eta_{ab} = (\eta_{\al \beta}, \eta_{AB} = \eta_{\hat \al \hat \beta} \dlt_{\ha, \hb}) ~.
\eea
Notice that the above construction ensures that the set of solutions (\ref{xi-sol}) is worth $(n-1)$ number of tangent vectors of $M$ as expected from (\ref{xi-bar-eq-M^n}). The latter can be explicitly checked to be satisfied by our construction.
\bea
\sum_p \dlt^{\al}{}_{\al_p} \xi_p^{\al_p}(x) &=& \sum_p \dlt^{\al}{}_{\al_p} O_{\ha p} \eta_{\hat \al}{}^{\beta_p} E_{(\beta_p)}{}^{\al_p}(x) = \lt( \sum_p O_{\ha p}\rt) E_{(\hat \al)}{}^{\al}(x) ~, \cr
&=& \sqrt{n} \lt( O^T O \rt)_{0 \ha} E_{(\hat \al)}{}^{\al}(x) = 0 ~.
\eea

We now show that the present example is a non-trivial special case of our general construction. Alhough $M^n$ is constructed by taking Cartesian product, because of the curvature of $M$ the internal frame of the a priori system cannot be aligned simply by the constant rotation in (\ref{Lambda-M^n}). This is demonstrated by computing
$\lambda^a{}_{ \bar b}$ in eq.(\ref{lambda-exp}) and showing that it is non-trivial. The rules of computation laid out in \S \ref{ss:M-data} enables one to find this expansion in terms of $M$-data. While the expansion for $\lambda^A{}_{\bar b}$ with 
$\bar b = \beta_p$ is given by,
\bea
\lambda^A{}_{\bar b} &=&  O_{\ha p} \lt[ \dlt^{\hat \al }{}_{\beta_p } - \Omega_{\gamma_1(\beta_p )}{}^{(\hat \al)} \xi_p^{\gamma_1} - {1\over 2} \Omega_{\gamma_1 \gamma_2 (\beta_p )}{}^{(\hat \al )}   \xi_p^{\gamma_1 \gamma_2}  
- {1\over 3!} \Omega_{\gamma_1 \gamma_2 \gamma_3 (\beta_p)}{}^{(\hat \al )} \xi_p^{\gamma_1 \gamma_2 \gamma_3} - \cdots \rt] ~, \cr && 
\label{lambda-exp-M^n}
\eea
the same for $\lambda^{\al}{}_{\bar b}$ is simply given by the above expression with the indices $\ha$ and $\hat \al$ replaced by $0$ and $\al$ respectively. The multi-indexed $\Omg$ coefficients can be read off from (\ref{multi-omega}) by using our rules in (\ref{tbar-prime-T}) and noting that the Weyl-weight of connection one form is zero. 
\bea
\Omega_{\gamma_1 \cdots \gamma_n }{}^{(\al )}{}_{(\beta )} &=& {1\over n} \lt[ 
\nabla^{t}_{\gamma_1} \Omg_{\gamma_2 \cdots \gamma_n }{}^{(\al)}{}_{(\beta)} 
- \Omg_{\gamma_1}{}^{(\al)}{}_{(\dlt)} \Omg_{\gamma_2 \cdots \gamma_n }{}^{(\dlt)}{}_{(\beta)} + \cdots   \rt] 
\eea
Here $\nabla^{t}$ is the total covariant derivative which annihilates the induced vielbein in $M$. The ellipses contain other terms required for symmetrization of coordinate indices. This clearly shows that the $\lambda$-matrix non-trivially changes as we move away from the submanifold. The expansion trivialises when $M$ is flat.

\subsubsection{How the method of \cite{cut-off} works }
\label{ss:cut-off}

Given that $\Dlt \hookrightarrow M^n$ is a non-trivial example, one may wonder how the relevant tubular geometry could be described in \cite{cut-off} without using this general construction. Our goal here is to explain this point in detail. 

There are two issues involved: (1) the indirect method itself which gives the final result for the expansion and (2) the verification, which explains how the result makes sense. It is easy to argue that the indirect method and the final results of \cite{cut-off} are correct. This is because the $\Lambda$-matrix is constant over the submanifold and therefore our rules for indirect method laid out in this work simply match with those used in \cite{cut-off}. However, because $\lambda$-matrix in non-trivial, our general construction non-trivially applies here and therefore the issue of verification (of eq.(\ref{ehat-e})) is not that simple, as we explain below. 

The key point is the following. Although the $\Lambda$-rotated internal a priori frame is not aligned away from the submanifold, the use of $\lambda$-matrix in the verification process has been avoided by using a trick that exists because of certain specificity of the problem. This happens to be related to the coordinate transformation that was explicitly constructed in \cite{cut-off} to all orders. Below we first recall this construction and then explain the point. 

The FNC $\hat z^a = (x^{\al}, \hat y^A)$ and the a priori system $\bar z^{\bar a}$ in (\ref{zbar-M^n}) were related by the following series of coordinate transformations,
\bea
\hat z^a \to z'^a = (x^{\al}, y'^A) : &&  y'^A =  \underline{ e'_{(B)}{}^A}   \hat y^B  ~, \cr
z'^a \to z^a = (x^{\al}, u^A) : &&  u^A = y'^A - \sum_{n\geq 0} {1\over (n+2)!} \underline{\gamma^A_{B^1 \cdots B^{n+2}}} y'^{B^1} \cdots y'^{B^{n+2}} ~, \cr
z^a \to \tilde z^a =(\tilde x^{\al}, u^A) : && \tilde x^{\alpha} = \tilde{\exp}_{\parallel}^{\alpha}(x, \tilde{\log}_{\perp} (x, u)) ~, \cr
\tilde z^a \to \bar z^{\bar a} : && \bar z^{\bar a} = \sqrt{n} (\Lambda^{-1})^{\bar a}{}_a \tilde z^a ~,
\label{coordinates-M^n}
\eea
where we have used the following notations. $e'^{(a)}{}_b$ is the vielbein in $z'$-system, $\gamma^a_{bcd\cdots}$ are the multi-indexed gamma coefficients in $z$-system defined in the same way as in (\ref{multi-gamma}). Finally, $\tilde{\exp}_{\parallel}$ is defined in the following way,
\bea
\tilde{\exp}_{\parallel}^{\alpha}(x, \tilde \xi) &=& x^{\alpha} - \sum_{n\geq 0} {1\over (n+2)!} 
\underline{\tilde \gamma^{\alpha}_{B^1 \cdots B^{n+2}}} \tilde \xi^{B^1} \cdots \tilde \xi^{B^{n+2}} ~,
\label{exp-tilde-alpha}
\eea
and $\tilde \xi = \tilde{\log}_{\perp}(x, u)$ is the inverse of the following map,
\bea
u^A &=& \tilde \xi^A - \sum_{n\geq 0} {1\over (n+2)!} \underline{\tilde \gamma^A_{B^1 \cdots B^{n+2}}} \tilde \xi^{B^1} \cdots \tilde \xi^{B^{n+2}} ~,
\eea 
where $\tilde \gamma^a_{bcd\cdots}$ are the multi-indexed gamma coefficients in $\tilde z$-system. 

As it was shown in \cite{cut-off}, the net transformation, when evaluated in terms of $M$-data, takes the following form\footnote{As explained in \cite{cut-off}, this result is also expected by construction.},
\bea
x_p^{\al_p} &=& x^{\al_p} + \Exp_x^{\al_p} (\xi_p) ~, 
\eea 
where $\Exp_x^{\al_p} (\xi_p)$ is the exponential map of $M$ around $x\in M$,
\bea
\Exp_x^{\al_p}(\xi_p) &=& \xi_p^{\al_p} - \sum_{n\geq 0} {1\over (n+2)!} \Gamma^{\al_p}_{\beta_p^1 \cdots \beta_p^{n+2}} (x)\xi_p^{\beta_p^1 \cdots \beta_p^{n+2}} ~,
\eea
$\Gamma^{\al}_{\beta \gamma \dlt \cdots}$ being the multi-indexed gamma coefficients (\ref{multi-gamma}) of $M$. 

It is a simple exercise to check that our coordinate transformation in (\ref{z-zhat}, \ref{exp}) (with suitable change of notation as adopted in this subsection), which is valid in general, also reproduces the same result when specialized to the present example and the rules of \S \ref{ss:M-data} are applied to express everything in terms of $M$-data.

Although (\ref{coordinates-M^n}) achieves the same results in several more steps, the advantage of it is the construction of 
$\tilde z$, which allows one to avoid the use of the non-trivial $\lambda$-matrix in the process of verification. This happens in the following way. Notice that (\ref{coordinates-M^n}) specifies only the coordinates, and it does not say anything regarding the internal frames. Implicit in this is the fact that starting from FNC, which is by construction aligned, all the other intermediate coordinates, namely $z' \to z \to \tilde z$ are also automatically aligned unless an additional internal rotation is imposed (which is neither necessary, nor has been done). Because of this the following identities, which have been used in \cite{cut-off}, hold,
\bea
\hat y^C \underline{\hat \omg_C{}^{(a)}{}_{(b)} } &=& \xi'^C \underline{\omg'_C{}^{(a)}{}_{(b)} } = \xi^C \underline{ \omg_C{}^{(a)}{}_{(b)} } = \tilde \xi^C \underline{\tilde \omg_C{}^{(a)}{}_{(b)} } = 0 ~, \cr
\hat y^{C_1 C_2 \cdots } \underline{\hat \omg_{C_1 C_2 \cdots}{}^{(a)}{}_{(b)} } &=& \cdots = \tilde \xi^{C_1 C_2 \cdots } \underline{\tilde \omg_{C_1 C_2 \cdots}{}^{(a)}{}_{(b)} } = 0~,
\label{aligned-ids}
\eea
where,
\bea
\tilde \xi^C = \xi^C = \xi'^C = \underline{ e_{(B)}{}^A } \hat y^B ~,
\eea
are the components of the same transverse vector in (\ref{xibar-M^n}) in different systems. This, however, is not true for the a priori system $\bar z$, as has been argued before. Therefore, though in the last step of (\ref{coordinates-M^n}), $\tilde z$ and $\bar z$ are related by the constant $\Lambda$-matrix, the internal frames are subject to a transformation given by the $\lambda$-matrix in (\ref{lambda-exp-M^n}). One can explicitly check, using the rules of \S \ref{ss:M-data}, that the equations in (\ref{aligned-ids}) do not hold in a priori system, which is why $\lambda$ is non-trivial in the first place.

What does the above feature have to do with the issue of verification? What was done in \cite{cut-off} is to first verify the following equation,
\bea
\hat e^{(a)}{}_b &=& \lt( \del z'^{b_1} \over \del \hat z^{b} \rt) \lt( \del z^{b_2} \over \del z'^{b_1} \rt) 
\lt( \del \tilde z^c \over \del z^{b_2} \rt) \tilde e^{(a)}{}_c ~. 
\eea
Various terms from the ordinary Taylor expansion of the RHS conspire to produce coefficients with the correct tensor structure at each order, without having to face the issue of a non-trivial $\lambda$-matrix as all the associated frames are aligned. Once these coefficients are obtained, they are related to the a priori system and subsequently re-expressed in terms of $M$-data - a process that involves only the constant $\Lambda$-matrix, given the fact that all the coefficients are evaluated on the submanifold. On the other hand, for the LHS, one evaluates the known tubular expansion coefficients in terms of $M$-data using the indirect method as usual. The work of \cite{cut-off} verifies that these two computations give the same results.

\subsection{KC background}
\label{s:KC}

The metric in the a priori coordinate system $\zb^{ a} = (z^0,  z_{\perp})$, $ z_{\perp} = \{ z^i \}~, (i = 1, 2, \cdots , (d-1))$ is given in matrix form as,
\bea
 g &=& \begin{pmatrix} 
-(1- \phi(\zb_{\perp})) & 0 \cr 0 & \dlt_{ij} + {k {\hr}^2 \over (1-k {\hr}^2) } \al_{ij} 
\end{pmatrix} ~, 
\label{gbar-kc}
\eea
where,
\bea
{\hr}^2 = \sum_i (z^i)^2 ~, \quad 
\al_{ij\cdots } = \al_i \al_j \cdots ~, \quad \al_i = \dlt_{ij}\al^j~, \quad \al^i = {z^i \over {\hr}} ~,
\eea
and $\phi (  z_{\perp} )$ satisfies the following condition, 
\bea
\phi (0) = 0~, \quad \del_i \phi (0) = 0~.
\eea
The $1$-dimensional submanifold considered is given by the following equations,
\bea
f^0 = x, \quad f^i = 0~,
\label{f-kc}
\eea
so that $x$ is an internal parametrization of the worldline. 

It was shown in \cite{kc-exact} that FNC $\hat z^{\hat a} = (x, \hat y^A)$, $(A=1,2, \cdots (d-1))$ can be exactly constructed and therefore the metric $\hat g_{ab}$ can be given in exact form. We summarize the results below. For $k=a^2 >0$, the exact coordinate transformation is given by,
\bea
z^0 = x~, && z^i = \al^i {\sin(a \hat{\hr}) \over a} ~,
\label{z-zhat-kc1}
\eea
where,
\bea
\hat{\hr}^2 &=& \sum_A (\hat y^A)^2 ~.
\eea
Note that, ${\hr} = {\sin(a\hat{\hr}) \over a}$ and $\al^i = {z^i \over {\hr}} = \dlt^i{}_A {\hat y^A \over \hat{\hr}} $.
The exact metric in FNC is given by,
\bea
\hat g &=& 
\begin{pmatrix}
-(1-\phi) & 0 \cr
0 & \al_{AB} + {\sin^2(a\hat{\hr}) \over a^2 \hat{\hr}^2}  (\dlt_{AB} - \al_{AB})  
\end{pmatrix} ~.
\label{ghat-kc}
\eea
Here, $\phi = \phi(  z_{\perp} (\hat y))$ is of course understood to be a function of $\hat y$. For $k=-a^2 <0$, the above results are modified by replacing sine functions by the corresponding hyperbolic functions. For our discussion below we shall specifically consider $k>0$, but generalization of the arguments to $k<0$ is straightforward. 

Our goal here is to recover the above results as a special case of our general construction. Noticing from (\ref{gbar-kc}) that $\underline{ g_{ a  b}} = \eta_{ a  b}$ and therefore, $\underline{ g_{ a  b}} = \dlt^{\hat a}{}_{ a} \dlt^{\hat b}{}_{ b} 
\underline{\hat g_{\hat a \hat b}}$, one concludes that,
\bea
K &=& \one_d ~.
\eea
This is enough information to compute the coordinate transformation (\ref{z-zhat}), which can be rewritten as an $\hat{\hr}$-expansion in the following way,
\bea
z^0 = x ~, \quad 
z^i = \al^i \hat{\hr} - \sum_{n\geq 0} {1\over (n+2)!} \underline{ \gamma^i_{j_1 j_2 \cdots j_{n+2}}} \al^{j_1 j_2 \cdots j_{n+2}} \hat{\hr}^{n+2} ~, 
\label{z-zhat-kc2}
\eea
where we have used,
\bea
\underline{ \gamma^0_{j_1 j_2 \cdots j_{n+2}}} &=& 0~,
\label{multigamma-0-kc}
\eea
which we have argued for in Appendix \ref{a:KC}. In order for (\ref{z-zhat-kc2}) to match with (\ref{z-zhat-kc1}), the following results need to be satisfied,
\bea
\underline{ \gamma^i_{j_1 j_2 \cdots j_{2n}}}  &=& 0 ~, \cr
\underline{ \gamma^i_{j_1 j_2 \cdots j_{2n+1}}} \al^{j_1 j_2 \cdots j_{2n+1}}  &=& (-1)^{n+1} k^n \al^i  ~.
\label{multigamma-i-kc}
\eea 
While we have argued for the first equation for any $n\geq 1$, we have checked the second equation with explicit computation up to $n=3$. 

In order to compute the metric $\hat g_{ab}$, we consider vielbein, for which we know the all order result. For our computation, we shall not need an explicit form of vielbein in a priori system. It will be enough for us to note that (1) it is time independent and (2) it is aligned on the submanifold, i.e. $\underline{ e^{( a)}{}_{ b}} = \eta^{ a}{}_{ b} $. The latter implies that,
\bea
\Lambda &=& \one_d ~.
\eea
Therefore, like in the previous example, every term, including the ones involving spin connection, transforms like a tensor. 
The expressions in (\ref{yhat-D-rho-result}, \ref{omg-hat-result}) work out to give,
\bea
\underline{ [(\hat y .\hat D^{t} )^s \hat \rho (\hat y)] } 
&=& \hat{\hr}^{s+2} \al^{i_1 \cdots i_s kl} 
\begin{pmatrix}
\underline{ D_{i_1} \cdots  D_{i_s}  r^0{}_{kl 0}} & 0 \cr 
0 & \underline{ D_{i_1} \cdots  D_{i_s}  r^i{}_{kl j} }
\end{pmatrix} ~, 
\label{yhat-D-rho-result-kc}  \\
\eh_0^{(a)}{}_b &=& \dlt^a{}_b ~.
\label{ehat-0-result-kc}
\eea
where we have used $\underline{  \omg_{0  a k}} = 0$ and the fact that $ r^0{}_{klj}$ and $ r^i{}_{kl0}$ vanish identically (see eqs.(\ref{riemann-kc})). We shall now compute the longitudinal and transverse components of vielbein in FNC separately. 

We begin by longitudinal components which will be computed up to quartic order. Using the results of Appendix \ref{a:theorem} one finds, 
\bea
\hat e^{(0)}{}_0 &=& 1 + { 1\over 2 }  \al^{i_1 i_2 } \underline{  r^0{}_{i_1 i_2 0} } \hat \hr^2 
+ { 1 \over 6 } \al^{i_1 i_2 i_3} \underline{  D_{i_1}  r^0{}_{i_2 i_3 0} } \hat \hr^3  
+ {1\over 24} \al^{i_1 \cdots i_4} \lt( \underline{  D_{i_1 i_2 }  r^0{}_{i_3 i_4 0} } 
+ \underline{  r^0{}_{i_1i_2 0 }  r^0{}_{i_3 i_4 0 } } \rt) \hat \hr^4  \cr
&& + O(\hat \hr^5) ~.
\eea
The above coefficients are computed using the results of Appendix \ref{a:KC}. These are given in terms of $ z_{\perp}$-derivatives of $\phi$. Therefore, it is more preferable to write the above expression as a power series in $z^i$ by using the second relation in (\ref{z-zhat-kc1}). The resulting expansion is given by,
\bea
\hat e^{(0)}{}_0 &=& 1 + { 1\over 2 }  \al^{i_1 i_2 } \underline{  r^0{}_{i_1 i_2 0} }  \hr^2 
+ { 1 \over 6 } \al^{i_1 i_2 i_3} \underline{  D_{i_1}  r^0{}_{i_2 i_3 0} }  \hr^3  \cr
&&
+ \lt[ {1\over 24} \al^{i_1 \cdots i_4} \lt( \underline{  D_{i_1 i_2 }  r^0{}_{i_3 i_4 0} } 
+ \underline{  r^0{}_{i_1i_2 0 }  r^0{}_{i_3 i_4 0 } } \rt) 
+ {k \over 6}  \al^{i_1 i_2 } \underline{  r^0{}_{i_1 i_2 0} } \rt]  \hr^4 
+ O( \hr^5) ~,
\eea
which, after we use results in (\ref{curvature-derivatives}), takes the following form,
\bea
\hat e^{(0)}{}_0 
&=& 1 - {1\over 4} \underline{\del_{i_1} \del_{i_2} \phi } z^{i_1 i_2}   
- {1\over 12} \underline{ \del_{i_1} \del_{i_2} \del_{i_3} \phi } z^{i_1 i_2 i_3}  \cr
&& - \lt( {1\over 48 } \underline{ \del_{i_1} \del_{i_2} \del_{i_3} \del_{i_4} \phi } 
+ {1\over 32} \underline{ \del_{i_1} \del_{i_2} \phi \del_{i_3} \del_{i_4} \phi } \rt) z^{i_1 \cdots i_4 } + O(z^5) ~, \cr
& \to & (1-\phi)^{1/2} ~, 
\label{ehat-long}
\eea
where in the last line we have indicated the functional form for which the expansion matches up to quartic order. Given the metric in (\ref{ghat-kc}), this is the expected result. 

Recall that \cite{tubular} provides tubular expansion coefficients of the vielbein components to all orders. As mentioned earlier, the result for the transverse components matches with the Riemann normal expansion as expected. However, the result for the longitudinal components is the non-trivial new result of \cite{tubular}. Note that the above computation provides a non-trivial check for this new result for the spin connection independent coefficients. 

For completeness we also derive the expansion for the transverse components to all orders. To this end we first note that the last equation in (\ref{curvature-derivatives}) implies: $\underline{ D_{i_1} \cdots  D_{i_n}  r^i{}_{jkl} } = 0 $, $\forall n\geq 1$ \footnote{This fact was first observed in \cite{msc}. }. 
Using this result in the first equation of (\ref{ehat-n-exp}) one gets,
\bea
\hat e^{(A)}{}_B &=& \dlt^A{}_B + \sum_{n \geq1} {1\over (2n+1)!} \hat \hr^{2n} \dlt^A{}_i \dlt_B{}^j \al^{i_1j_1i_2j_2 \cdots i_nj_n} 
\underline{\rb^i{}_{i_1j_1k_1} \rb^{k_1}{}_{i_2 j_2 k_2} \cdots \rb^{k_{n-1}}{}_{i_n j_n j} } ~, \cr
&&
\eea
which, under further manipulations, produces the following result,
\bea
\hat e^{(A)}{}_B &=& \al^A{}_B + {\sin (a\hat \hr) \over a\hat \hr} (\dlt^A{}_B - \al^A{}_B) ~.
\label{ehat-trans}
\eea
The results in (\ref{ehat-long}, \ref{ehat-trans}) give rise to the metric in (\ref{ghat-kc}).

\section{Outlook}
\label{end}

Although the result of \cite{tubular} specifies all the tubular expansion coefficients of vielbein with their tensor characters manifest, in general it is not directly usable in a given problem. This is simply because a physics problem is usually specified in a certain convenient a priori coordinate system. Our work specifies the expansion directly in terms of the a priori data. Therefore the result is readily usable in any given situation, as has been demonstrated through a couple of examples. 

Our general construction explicitly relates coordinate and non-coordinate frames associated to a priori system and the Fermi system. The coordinate invariant relation between the non-coordinate frames, given by the $\lambda$-matrix, can be interpreted to be the analogue of the derivative of exponential map which relates the coordinate frames. We introduce a notion of alignment of the internal frames according to which the expansion for $\lambda$ trivializes when the a priori frame is aligned. In other words, the Fermi system and all other systems that are related to it by pure coordinate transformation (without altering the internal frames) are all aligned. Another way of seeing this is as follows. The coordinate conditions for Fermi system involves spin connection. This gives rise to an infinite number of conditions on the derivatives of spin connection at the submanifold. Analogous conditions, in covariant form, are also applicable for all those systems which are obtained from FNC by pure coordinate transformations. These are precisely the triviality conditions for $\lambda$.

The case of $\Dlt \hookrightarrow M^n$ is a non-trivial example of our general construction in the sense that the corresponding 
$\lambda$ is non-trivial. It then appears to be a puzzle how the indirect method in \cite{cut-off} was verified without the prior knowledge of our present construction which uses the expansion for $\lambda$ in a crucial manner. We explain this in detail where the aforementioned observations regarding alignment play a crucial role. 

The KC background \cite{kc-exact} proves to be useful as a demonstrative example of our general construction. However, all the second class spin connection terms in the vielbein expansion vanish in this specific example. This is typically the case for those examples where FNC is found exactly. Such terms, on the other hand, are expected to be important in, for example, the general formulation of large-$D$ black holes. This forces us to perform the cubic-order-verification. Although this is a long and tedious computation as there are nearly $100$ terms to manipulate, we manage to use certain tricks to control it. 

We end this section by spelling out a technical motivation for the present work besides the ones mentioned in \S \ref{motiv}. This is in the context of the tubular geometry of loop space $LM$. The latter was found in \cite{cut-off} using the tubular expansion theorem of \cite{tubular} which assumes uniqueness of the pair of points $\{P, Q\}$ as described in our general setup in \S \ref{ss:setup} and the geodesic connecting them. For $LM$, $P$ corresponds to a specific non-zero loop in $M$. In \cite{cut-off}, an independent definition of $Q \in M$ was given as the centre of mass (CM) of the loop which was used in the indirect method. The geodesic $(P,Q)$ in $LM$ and the definition of CM of the loop in $M$ must be compatible with each other in certain sense. Verification of these compatibility conditions has posed immense problem \cite{geodesic} due to the fact that the coordinate transformation (\ref{coordinates-M^n}) is given in multiple stages. Such compatibility conditions should be inbuilt in our general construction where the relevant coordinate transformation (\ref{z-zhat}, \ref{exp}) is given only in one step. We hope our present work will ease the computational challenge drastically.

\begin{center}
{\bf Acknowledgement}
\end{center}

The author would like to thank Sumit R. Das, Ghanashyam Date and Alfred D. Shapere for useful discussions. The kind support from Department of Physics and Astronomy, University of Kentucky, USA during the final stage of the work is gratefully acknowledged. This work was partially supported by National Science Foundation grant NSF-PHY/1521045. 

\appendix

\section{The tubular expansion theorem} 
\label{a:theorem}  

The tubular expansion theorem, as derived in \cite{tubular}, spells out, for an arbitrary submanifold embedding $M \hookrightarrow L$, all the tubular expansion coefficients of vielbein with manifest tensorial properties in closed form. Here we review this result. 

Given the geometric set up and notations as described in \S \ref{motiv} and \S \ref{general}, whenever the following coordinate conditions (for FNC) are satisfied within a tubular neighbourhood, 
\bea
\hat e^{(\hat a)}{}_B \hat y^B = \underline{ \hat e^{(\hat a)}{}_B } \hat y^B = \dlt^{\hat a}{}_B \hat y^B ~, && \quad \hat y^B \hat \omg_B{}^{(\hat a)}{}_{(\hat b)} = 0 ~, 
\label{FNC-cond}
\eea
($\hat \omg_{\hat a}$ being the components of spin connection in FNC), the following result holds,
\bea
\hat e^{(\hat a)}{}_{\hat b} &=& \sum_{n=0}^{\infty} \hat e_n^{(\hat a)}{}_{\hat b} ~, 
\eea
where $\hat e_n^{(\hat a)}{}_{\hat b}$ is the contribution at $n$-th order in curvature. This is given by\footnote{Note that in the previous versions of \cite{tubular}, the closed-form expression for the coefficients of the second class appeared incorrectly (originally due to a typo). This has been corrected in the latest version. },
\bea
\hat e_n^{(\hat a)}{}_B &=& \sum_{ \{s \} } {\cal F}^{(n)}_{\perp}(\{s\}) \underline{\hat \pi^{(\hat a)}{}_{(\hat c)} (\{s\} , \hat y) 
\hat e^{(\hat c)}{}_B } ~,  \cr
\hat e_n^{(\hat a)}{}_{\beta} &=& \sum_{ \{s\} } \lt[ {\cal F}^{(n)}_{\parallel} (\{s\} ) 
\underline{\hat \pi^{(\hat a)}{}_{(\hat c)} (\{s\}, \hat y) \hat e^{(\hat c)}{}_{\beta} } 
+ {\cal F}^{(n)}_{\perp}(\{s\})  \underline{\hat \pi^{(\hat a)}{}_{(\hat c)} (\{s\} , \hat y) \hat \omg_{\beta}{}^{(\hat c)}{}_C } \hat y^C \rt] ~, \cr
&& 
\label{ehat-n-exp}
\eea
where $\{ s\}$ is a set of $n$ integers $s_1, \cdots, s_n $, $s_i \geq 0$ and,
\bea
\cF_{\parallel}^{(n)}(\{s\}) &=& {C_{\parallel}^{(n)}( \{s\} ) \over (s_1+ s_2+\cdots + s_n + 2n)!} ~, \cr
\cF_{\perp}^{(n)}(\{s\}) &=& {C_{\perp}^{(n)}( \{ s \} ) \over (s_1 + s_2 + \cdots + s_n + 2n +1)!}~,
\label{cF1cF3} \\
&& \cr
C_{\parallel}^{(n)}( \{ s \} ) &=& C^{s_1+s_2\cdots + s_n + 2n -2}_{s_1} C^{s_2+s_3+\cdots + s_n + 2n-4}_{s_2}  \cdots 1~, \cr
C_{\perp}^{(n)}( \{ s \} ) &=& C^{s_1+s_2\cdots + s_n + 2n -1}_{s_1} C^{s_2+s_3+\cdots + s_n + 2n-3}_{s_2}  \cdots C^{s_n+1}_{s_n} ~,
\label{C1C3}
\eea
where $C^n_r$ are binomial coefficients. Furthermore\footnote{We follow the same definition of Riemann curvature as in \cite{nakahara}. } , 
\bea
\underline{ \hat \pi (\{s\}, \hat y)} &=& \underline{(\hat y .\hat D^{t} )^{s_1} \hat \rho (\hat y)} \cdots 
\underline{(\hat y .\hat D^{t} )^{s_n} \hat \rho (\hat y)}~, \cr
\underline{ [(\hat y .\hat D^{t} )^s \hat \rho (\hat y)]^{(\hat a)}{}_{(\hat b)} } &=& \hat y^{C_1 \cdots C_s D E} \underline{\hat D^{t}_{C_1} \cdots \hat D^{t}_{C_s} \hat r^{(\hat a)}{}_{DE (\hat b)} } ~, \cr
&=& \hat y^{C_1 \cdots C_s DE} \underline{\hat \del_{C_1} \cdots \hat \del_{C_s} \hat r^{(\hat a)}{}_{DE (\hat b)} } ~. 
\label{pi-hat}
\eea
Finally,
\bea
\underline{\hat e^{(\hat a)}{}_{\beta} } = \dlt^{\hat a}{}_{\al} \underline{ \hat e^{(\al)}{}_{\beta} }  ~, \quad
\underline{\hat e^{(\hat a)}{}_B } = \dlt^{\hat a}{}_B ~.
\label{underline-e}
\eea
Note that the equality between the second and third lines of (\ref{pi-hat}) can be established by using the following result,
\bea
\hat y^{C_1 \cdots C_s} \hat \del_{C_1} \cdots \hat \del_{C_{s-1}} \hat \omg_{C_s}{}^{(\hat a)}{}_{(\hat b)} &=& 0~,
\label{multi-der-omg}
\eea
which follows from the second equation in (\ref{FNC-cond}).

The above results take the following form in terms of a priori data,
\bea
\underline{ [(\hat y .\hat D^{t} )^s \hat \rho (\hat y)]^{(\hat a)}{}_{(\hat b)} } 
&=& \hat y^{A_1 \cdots A_s DE} \eta^{\hat a \hat a'} \Lambda^{ a}_{\hat a'}{}^{ a_1}_{A_1}{}^{\cdots}_{\cdots}{}^{ a_s}_{A_s}{}^{ d}_D{}^{ e}_E 
\underline{ D^{t}_{( a_1)} \cdots  D^{t}_{( a_s)}  r_{( a  d  e b)} } ~, 
\label{yhat-D-rho-result} 
\cr && \\
\underline{ \hat \omg_{\beta}{}^{(\hat a)}{}_C } \hat y^C 
&=& \eta^{\hat a \hat a'}  \lt( f^{ b}{}_{, \beta} \Lambda_{\hat a'}^{ a}{}^c_C \underline{  \omega_{ b ( a c)}} 
+  \eta_{ac} \Lambda{}_{ \hat a' }^a{}_{\beta }{}_C^{ c} \rt) \hat y^C ~, 
\label{omg-hat-result} 
\eea
where we have used the following notations,
\bea
\Lambda^{ a}_{\hat a}{}^{ b}_{\hat b}{}^{\cdots}_{\cdots} := \Lambda_{\hat a}{}^{ a} \Lambda_{\hat b}{}^{ b} \cdots ~, \quad  D^{t}_{( a)} :=  e_{( a)}{}^{ b}  D^{t}_{ b} ~. 
\label{notations}
\eea

\section{Computation of metric-expansion }
\label{a:comp}

Here we give the details of the computation of metric expansion in \S \ref{s:metric}. In any given term in the expansion of vielbein in FNC, the spin connection appears at most linearly with its last index contracted with $\hat y$. Therefore, in the expansion of metric, any term is at most quadratic in spin connection with the last index of each factor being contracted with a factor of $\hat y$. Moreover, the first index of each factor of the spin connection must be free and therefore should match with one of the indices of the metric. Following are the possible structure of such terms (keeping in mind if the first index of spin connection (in FNC) is transverse, then it vanishes. See eq.(\ref{omg-hat-sub})) and the corresponding results,
\bea
\hat l_{1 \al \beta} &=& \underline{\hat \omega_{\al \beta C} } \hat y^C = \lt( f^{ a}{}_{, \al} f^{ b}{}_{, \beta} \Lambda_C{}^{ c} \underline{ \omega_{ a  b ( c)}  } + f^{ b}{}_{, \beta} \del_{\al}  \Lambda_C{}^{ c } \underline{  e_{( c )  b} }  \rt) \hat y^C   ~, \cr
\hat l_{1 \al B} &=& \underline{\hat \omega_{\al B C} } \hat y^C = \lt( f^{ a}{}_{, \al} \Lambda^{ b}_B{}^{ c}_C 
\underline{ \omega_{ a ( b  c) }} 
+ \Lambda^{ b}_B{}_{\al}{}^{ c}_C \eta_{ b  c} \rt) \hat y^C ~,  \cr
\hat l_{2 \al \beta} &=& \underline{\hat \omega_{\al}{}^{(\hat e)}{}_C \hat t_{(\hat e) \beta }} \hat y^C 
= \lt( f^{ a}{}_{, \al} f^{ b}{}_{, \beta} \Lambda_C{}^{ c} \underline{  \omega_{ a}{}^{( e)}{}_{( c)}  t_{( e)  b} } + \del_{\al} \Lambda_C{}^{ c}
f^{ b}{}_{, \beta}  \underline{  t_{( c)  b} } \rt) \hat y^C 
~, \cr
\hat l_{2 \al B} &=& \underline{ \hat \omega_{\al}{}^{(\hat e)}{}_C \hat t_{(\hat e) B} } \hat y^C = \lt( f^{ a}{}_{, \al} \Lambda^{ b}_B{}^{ c}_C 
\underline{  \omega_{ a}{}^{(e)}{}_{( c)}  t_{(  b  e) } } 
+ \Lambda^{ b}_B{}_{\al}{}^{ c}_C \underline{  t_{(  b  c)} } \rt) \hat y^C ~, \cr
\hat q_{1 \al \beta} &=& \underline{ \hat \omega_{\al}{}^{(\hat e)}{}_{C_1} \hat \omega_{\beta (\hat e) C_2 }} \hat y^{C_1 C_2} ~, \cr
&=& \lt[ f^{ a}{}_{, \al} f^{ b}{}_{, \beta } \Lambda^{ c_1}_{C_1}{}^{ c_2}_{C_2} 
\underline{  \omega_{ a}{}^{( e)}{}_{( c_1) }  \omega_{ b}{}_{( e  c_2 ) } }  
- \lt( f^{ a}{}_{, \al} \Lambda^{ c_1}_{C_1}{}_{\beta }{}^{ c_2}_{C_2} \underline{  \omega_{ a (  c_1  c_2) } } + \al \leftrightarrow \beta \rt) 
+ \Lambda_{\al}{}^{ c_1}_{C_1}{}_{\beta }{}^{ c_2}_{C_2} \eta_{ c_1  c_2} \rt]  \hat y^{C_1 C_2} ~, \cr
\hat q_{2 \al \beta} &=& \underline{ \hat \omega_{\al}{}^{(\hat e)}{}_{C_1} \hat \omega_{\beta }{}^{(\hat f)}{}_{C_2} \hat t_{(\hat e \hat f)} } \hat y^{C_1 C_2} ~, \cr
&=& \lt[ f^{ a}{}_{, \al} f^{ b}{}_{, \beta } \Lambda^{ c_1}_{C_1}{}^{ c_2}_{C_2}  
\underline{ \omega_{ a}{}^{( e)}{}_{( c_1)}  \omega_{ b}{}^{( f)}{}_{( c_2)}  t_{( e  f)} }  
+ \lt( f^{ a}{}_{, \al} \Lambda^{ c_1}_{C_1}{}_{\beta}{}^{ c_2}_{C_2}   
\underline{ \omega_{ a}{}^{( e)}{}_{( c_1)}  t_{( e  c_2 )} }  + \al \leftrightarrow \beta \rt) \rt. \cr
&& \lt. + \Lambda_{\al}{}^{ c_1}_{C_1}{}_{\beta}{}^{ c_2}_{C_2}   \underline{  t_{( c_1  c_2 )} } \rt] \hat y^{C_1 C_2} ~.
\label{forms}
\eea
where $\hat t_{(\hat e) \hat b}$ and $\hat t_{(\hat e \hat f)} = \hat t_{(\hat f \hat e)}$ are tensors. To compute the above forms we have used the following transformation law,
\bea
\hat \omega_{\hat a}{}^{(\hat b)}{}_{(\hat c)} 
&=& k^{ a}{}_{\hat a} \lambda^{\hat b}{}_{ b} \lambda_{\hat c}{}^{ c}  \omega_{ a}{}^{( b)}{}_{ (c)} 
+ \lambda^{\hat b b} \del_a \lambda_{\hat c b}   ~.
\eea

Up to quartic order we get the following results from \cite{tubular},
\bea
\hat g_{\al \beta} 
&=&  \underline{ G_{\al \beta} } + (\underline{\hat \omg_{\al \beta C}} + \underline{\hat \omg_{\beta \al C} } ) \hat y^C 
+ ( \underline{\hat r_{\al C_1C_2 \beta} } 
+ \underline{\hat \omg_{\al}{}^{\hat a}{}_{C_1} \hat \omg_{\beta \hat a C_2} } ) \hat y^{C_1 C_2}   \cr
&&
+ \lt\{ 
{1\over 3 } \underline{ \hat D^{t}_{C_1} \hat r_{\al C_2 C_3 \beta} } 
+ {2\over 3} (\underline{\hat r_{\al C_1C_2 \hat a } \hat \omg_{\beta}{}^{\hat a}{}_{C_3} } + \al \leftrightarrow \beta )
\rt\}  \hat y^{C_1C_2C_3}  \cr
&& 
+ \lt\{ {1 \over 12} \underline{ \hat D^{t}_{C_1} \hat D^{t}_{ C_2} \hat r_{\al C_3C_4 \beta} }
+ {1\over 3} \underline{\hat r_{\al C_1C_2 \hat a } \hat r^{\hat a}{}_{ C_3 C_4 \beta} }
+  {1\over 4} (\underline{ \hat D^{t}_{C_1 } \hat r_{\al C_2 C_3 \hat a } \hat \omg_{\beta}{}^{\hat a}{}_{C_4} }  
+ \al \leftrightarrow \beta )  \rt. \cr
&& \lt. 
+ {1\over 3}  \underline{\hat r_{\hat a C_1C_2 \hat b} \hat \omg_{\al}{}^{\hat a}{}_{C_3} \hat \omg_{\beta}{}^{\hat b}{}_{C_4} } \rt\} 
\hat y^{C_1 \cdots C_4}  ~,   
\label{ghat-al-beta-gen} \\
&& \cr
\hat g_{\al B } 
&=& \underline{\hat \omg_{\al B C} } \hat y^C 
+  {2 \over 3} \underline{ \hat r_{\al C_1 C_2 B} } \hat y^{C_1 C_2 } 
+ \lt( {1 \over 4 } \underline{ \hat D^{t}_{C_1} \hat r_{\al C_2 C_3 B} } 
+ {1 \over 3} \underline{\hat \omg_{\al }{}_{\hat a C_1} \hat r^{\hat a}{}_{ C_2 C_3 B} } \rt) \hat y^{C_1 C_2 C_3 }  \cr
&&
+ \lt( {1 \over 15} \underline{ \hat D^{t}_{C_1 C_2 } \hat r_{\al C_3 C_4 B} } 
+ {2 \over 15 } \underline{ \hat r_{\al C_1C_2 \hat a} \hat r^{\hat a}{}_{C_3 C_4 B } } 
+ {1 \over 6 } \underline{\hat \omg_{\al }{}_{\hat a C_1}  \hat D^{t}_{C_2} \hat r^{\hat a}{}_{C_3 C_4 B} } \rt) \hat y^{C_1 \cdots C_4 } ~,  \cr
&&
\label{ghat-al-B-gen} \\
\hat g_{AB} 
&=& \eta_{AB} + {1 \over 3} \underline{ \hat r_{A C_1 C_2 B} } \hat y^{C_1 C_2 } 
+ {1 \over 6 }  \underline{ \hat D^{t}_{C_1} \hat r_{A C_2 C_3 B} } \hat y^{C_1 C_2 C_3 }  \cr
&&
+ \lt( {1 \over 20}  \underline{ \hat D^{t}_{C_1 C_2 } \hat r_{A C_3 C_4 B} } 
+ {2 \over 45} \underline{ \hat r_{A C_1C_2 \hat b} \hat r^{\hat b}{}_{C_3 C_4 B } } \rt)  \hat y^{C_1 \cdots C_4 } ~,
\label{ghat-A-B-gen}
\eea
Here one can explicitly see that all the terms are of the forms described in eqs.(\ref{forms}). Using these expressions in  the above equations we get the results as given in eqs.(\ref{lhs-g-al-beta}, \ref{lhs-g-al-B}, \ref{lhs-g-A-B}).

\section{Computations for KC background}
\label{a:KC}

Here we give details of certain computations needed for the arguments presented in \S \ref{s:KC}. The basic ingredients are the non-zero components of the Christoffel symbols. These are calculated to be,
\bea
 \gamma^0_{0i} &=& - { \del_i \phi \over 2 (1- \phi) } ~, \cr
 \gamma^i_{00} &=& - {1\over 2} (\dlt^{i j} - k{\hr}^2 \al^{ij}) \del_j \phi ~, \cr
 \gamma^i_{jk} &=& { k^2 {\hr}^3 \over (1-k{\hr}^2) } \al^i_{jk} + k{\hr} \al^i \dlt_{jk}  ~,
\label{christoffel-kc}
\eea
where we use the notation: $\al^i_{jk} = \al^i \al_j \al_k$. We first consider how the coordinate transformation (\ref{z-zhat-kc2}), as given by our construction, reproduces the known result (\ref{z-zhat-kc1}). The first part of (\ref{z-zhat-kc2}) uses the result (\ref{multigamma-0-kc}). This directly follows from the above results for the Christoffel symbols. We now turn to the first part of (\ref{z-zhat-kc2}). The argument requires eqs.(\ref{multigamma-i-kc}) to be correct. To show this one may proceed as follows. Noticing from (\ref{christoffel-kc}),
\bea
\underline{ \gamma^{ i}_{ j  k} } &=& 0~,
\eea
one concludes that $\underline{ \gamma^i_{j_1 \cdots j_n} } $ is a sum of terms where each term is a product of factors, each of which being a higher derivative of the transverse Christoffel symbols given by the last equation in (\ref{christoffel-kc}). In order to compute such higher derivatives one may proceed as follows. One first considers the ordinary Taylor expansion of $ \gamma^i_{jk}$ and rewrites it in the following way,
\bea
 \gamma^i_{ j k } &=& \sum_{n\geq 0} {1\over n!} \underline{ \del^n_{j_1\cdots  j_n }  \gamma^i_{j k } } z^{j_1}
\cdots z^{ j_n} 
= \sum_{n\geq 0} {1\over n!} \al^{j_1\cdots j_n} \underline{ \del^n_{ j_1 \cdots j_n}  \gamma^i_{ j k } } {\hr}^n   ~,
\eea
such that the coefficient of $ \hr^n$ is given by,
\bea
\underline{ \gamma^i_{(n) jk} } &=& {1\over n!} \al^{j_1\cdots j_n} \underline{ \del^n_{ j_1 \cdots j_n}  \gamma^i_{ j k } } ~,\eea
By computing this coefficient from the last equation of (\ref{christoffel-kc}) and substituting it in the above equation one finds (for $n \geq 1$) ,
\bea
\underline{ \del_{j_1}  \gamma^i_{jk} } &=& k \al^i_{j_1} \dlt_{jk} ~, \cr
\underline{ \del^{2n}_{j_1 \cdots j_{2n}}  \gamma^i_{jk} } &=& 0 ~,  \cr 
\underline{ \del^{2n+1}_{j_1 \cdots j_{2n+1}}  \gamma^i_{jk} } &=& (2n+1)! k^{n+1} \al^i_{j_1 \cdots j_{2n+1} jk} ~.
\eea	
Because of the above results and the fact that $\al^i_i = 1$, one finds the following simplified answer, for $p \geq 1$ and 
$n_k \geq 0 $~, $ k = 1, \cdots, (p+1)$,  
\bea
\theta^i_{p+1}(n_1, \cdots , n_{p+1}) & \equiv & 
\underline{\lt[ (\del^{2n_1+1}  \gamma ) (\del^{2n_2+1}  \gamma) \cdots (\del^{2n_{p+1} +1}  \gamma ) \rt]^i_{j_1 \cdots j_m} } \al^{j_1 \cdots j_m}  \cr 
&=& \prod_{k=1}^{p+1} \lt\{ (2n_k+1)! k^{n_k+1} \rt\} \al^i ~, 
\label{gen-term-kc}
\cr
&&
\eea
where,  
\bea
m = 2 \displaystyle \sum_{k=1}^{p+1} n_k + 2p + 3 ~. 
\label{no-of-js}
\eea
$\theta^i_{p+1}(\{n \}) $ has $(p+1)$ factors of terms each of which is a higher derivative of transverse Christoffel symbol of odd order. Each such factor gives one upper index giving rise to a total of $(p+1)$ such indices. Out of these, only one, given by $i$, is un-contracted. All the others are contracted with $p$ lower indices coming from the derivatives and Christoffel symbols. The remaining un-contracted lower indices are $j_1, \cdots , j_m$. The above equation shows that all terms of the same general structure as described above but with different contractions give the same result.

Our remaining job is to figure out the linear combination in which the terms (\ref{gen-term-kc}) appear in $\underline{ \gamma^i_{j_1 j_2 \cdots j_n} } \al^{j_1 j_2 \cdots j_n} $. The restriction in (\ref{no-of-js}) establishes the first equation in (\ref{multigamma-i-kc}). For the LHS of the second equation, we have performed explicit counting to find the desired linear combinations for the first few cases. The results, which support (\ref{multigamma-i-kc}), are as follows,
\bea
\underline{ \gamma^i_{j_1 j_2 j_3} } \al^{j_1 j_2 j_3} &=& \theta^i_1(0) = k \al^i ~, \cr
\underline{ \gamma^i_{j_1 \cdots j_5 } } \al^{j_1 \cdots j_5} &=& \underline{ \del^3_{(j_1 j_2 j_3 }  \gamma_{ j_4 j_5)} }
- 4 \underline{ \del_{( j_1} \gamma^k_{j_2 j_3 } \del_{j_4}  \gamma^i_{ k j_5) } }
- 3 \underline{ \del_{( j_1 }   \gamma^k_{j_2 j_3} \del_k  \gamma^i_{j_4 j_5 ) } } ~,  \cr
&=& \theta^i_1(1) - 7 \theta^i_2 (0,0)  = (3! - 7) k^2 \al^i = - k^2 \al^i ~, \cr
\underline{ \gamma^i_{j_1 \cdots j_7 } } \al^{j_1 \cdots j_7 }  &=& \underline{ \del^5_{( j_1 \cdots j_5 } \gamma^i_{j_6 j_7 ) } }
- 36 \underline{ \del_{ (j_1 \cdots j_3} \gamma^k_{j_4 j_5 } \del_{j_6} \gamma^i_{k j_7 ) }  }
+ 92 \underline{ \del_{ (j_1} \gamma^k_{j_2 j_3 } \del_{j_4} \gamma^{k_1}_{j_5 j_6} \del_{j_7 ) } \gamma^i_{k_1 k } } \cr
&&
- 5 \underline{ \del_{(j_1 } \gamma^k_{j_2 j_3 } \gamma^i_{j_4 \cdots j_7 ) k} } ~, \cr
&=& \theta^i_1(2) - 36 \theta^i_2(1, 0) + 92 \theta^i_3(0, 0, 0) - 5 \theta^i_2(0, 1) + 35 \theta^i_3(0, 0, 0)   ~, \cr
&=& \theta^i_1(2) - 41 \theta^i_2(0, 1) + 92 \theta^i_3(0, 0, 0) + 35 \theta^i_3(0, 0, 0)   ~, \cr
&=& \lt( 5! - 41 \times 3! + 92 + 35 \rt) k^3 \al^i = k^3 \al^i ~.
\eea

We now proceed to compute, using the a priori system, the Riemann curvature tensor and its covariant derivatives that are needed to compute the metric-expansion. The only non-zero components are $\rb^0{}_{kl0}$, $\rb^i{}_{00l}$ and $\rb^i{}_{jkl}$. We need only the following results,
\bea
\rb^0{}_{kl0} 
&=& - { \del_k \phi \del_l \phi \over  4(1- \phi )^2} 
- { \del_k \del_l \phi \over  2(1- \phi )} 
+ {1\over 2} \lt\{ {k^2 {\hr}^3 \over (1- k{\hr}^2)} \al^i{}_{k l} 
+ k{\hr} \al^i \dlt_{ kl} \rt\} { \del_i \phi \over (1- \phi )}  ~, \cr
\rb^i{}_{jkl} &=& \dlt^i{}_k \lt[ {k^2 {\hr}^2 \over (1-k{\hr}^2)} \al_{j l} + k \dlt_{jl} \rt] - k \leftrightarrow l ~. 
\label{riemann-kc}
\eea
The following results are needed for the computation of \S \ref{s:KC}. 
\bea
\al^{i_1 kl} \underline{ D_{i_1}  r^0{}_{kl 0} } 
&=& - {1\over 2} \al^{i_1 kl} \underline{\del_{i_1} \del_k \del_l \phi} ~, \cr 
\al^{i_2 i_1 kl} \underline{ D_{i_2}  D_{i_1}  r^0{}_{kl 0} } &=& \al^{i_1 \cdots i_4 } 
\lt[ - \underline{ \del_{i_1} \del_{i_2} \phi \del_{i_3} \del_{i_4} \phi } 
- {1\over 2} \underline{ \del_{i_1} \del_{i_2} \del_{i_3} \del_{i_4} \phi }  \rt] 
+ 2 k \al^{i_1 i_2} \underline{ \del_{i_1} \del_{i_2} \phi} ~, \cr
\al^{i_1 \cdots i_4} \underline{ r^0{}_{i_1 i_2 0}  r^0{}_{i_3 i_4 0} } &=& {1\over 4 } \al^{i_1 \cdots i_4}  \underline{ \del_{i_1} \del_{i_2} \phi \del_{i_3} \del_{i_4} \phi } ~, \cr
 D_i  r^i{}_{klj} &=& 0~, 
\label{curvature-derivatives}
\eea

\section{Details of verification}
\label{a:verification}

Here we shall present the details of verification of eq.(\ref{ehat-e}) as described in \S \ref{verification}. For LHS of this equation, we need the following results from \cite{tubular} (see Appendix \ref{a:theorem}),
\bea
\hat e^{(\hat a)}{}_B 
&=& \dlt^{\hat a}{}_B + {1 \over 6} \underline{ \hat r^{(\hat a)}{}_{C_1 C_2 B} } \hat y^{C_1 C_2 } + O(\hat y^3) ~, \cr
\hat e^{(\hat a)}{}_{\beta} 
&=& \underline{\hat e^{(\hat a)}{}_{\beta} } + \underline{\hat \omg_{\beta}{}^{(\hat a)}{}_C} \hat y^C 
+ { 1\over 2 }  \underline{ \hat r^{(\hat a)}{}_{C_1 C_2 \beta} } \hat y^{C_1 C_2 } 
+ { 1 \over 6 } \lt( \underline{ \hat D^{t}_{C_1} \hat r^{(\hat a)}{}_{C_2 C_3 \beta} } 
+ \underline{ \hat r^{(\hat a)}{}_{C_1C_2 \hat d} \hat \omg_{\beta}{}^{\hat d}{}_{C_3}  } \rt) \hat y^{C_1 C_2C_3 } \cr
&& + O(\hat y^4) ~.  
\label{lhs-fnc}
\eea
Notice that the second equation is expanded up to cubic order. This cubic term is the one of interest for the test as described in \S \ref{s:cubic}. We rewrite the above results in terms of a priori data using our prescription,
\bea
\hat e^{(\hat a)}{}_B 
&=& \Lambda^{\hat a}{}_{ a} \Lambda_B{}^{ b} \lt[ \dlt^{ a}{}_{ b}  
+ {1 \over 6}  \underline{  r^{( a)}{}_{  c_1  c_2  (b) } }  \xi^{ c_1  c_2}  \rt] + O( \xi^3) ~, 
\label{lhs-apriori-B} \\
\hat e^{(\hat a)}{}_{\beta} 
&=& 
\Lambda^{\hat a}{}_{ a}  f^{ b}{}_{, \beta} \underline{  e^{( a)}{}_{ b} } 
+ \lt( \Lambda^{\hat a}{}_{ a} f^{ b}{}_{, \beta} \underline{  \omega_{ b}{}^{( a)}{}_{ c} } 
- \del_{\beta } \Lambda^{\hat a}{}_{ a } \underline{ e^{( a) }{}_{ c}}  \rt)  \xi^{ c} 
+ { 1\over 2 }  \Lambda^{\hat a}{}_{ a} f^{ b}{}_{, \beta} 
\underline{  r^{(  a)}{}_{  c_1  c_2  b } }  \xi^{ c_1  c_2 } \cr
&&
+ { 1 \over 6 } \lt\{ \Lambda^{\hat a}{}_{ a} f^{ b}{}_{, \beta} 
\underline{  D^{t}_{ c_1}  r^{( a)}{}_{  c_2  c_3  b } } 
+ \Lambda^{\hat a}{}_{ a} \underline{  r^{( a)}{}_{  c_1  c_2 ( d)} } 
\lt( f^{ b}{}_{, \beta} \underline{  \omega_{ b}{}^{( d)}{}_{ c_3 }} - \Lambda_{\hat d}{}^{ d}  \del_{\beta } \Lambda^{\hat d}{}_{e} \underline{ e^{( e)}{}_{c_3} }  \rt) \rt\}  \xi^{ c_1  c_2  c_3 }  \cr
&& + O( \xi^4)  ~.
\label{lhs-apriori-beta}
\eea

The ingredients needed to compute RHS of eq.(\ref{ehat-e}) are given by eq.(\ref{lambda-exp}) and,
\bea
k^{ a}{}_B 
&=& \lt[ \dlt^{ a}{}_{ b} 
- \underline{ \gamma^{ a}_{ b  c_1}}  \xi^{ c_1} 
- {1\over 2} \underline{ \gamma^{ a}_{ b  c_1  c_2 }}  \xi^{ c_1  c_2}  \rt] K^{ b}{}_B 
+ O( \xi^3)   ~, 
\label{k-exp-B}\\
k^{ a}{}_{\beta} 
&=& f^{ b}{}_{, \beta} \lt[ \dlt^{ a}{}_{ b} 
+ \lt( - \underline{ \gamma^{ a}_{ b  c_1} } 
+ \underline{ \omg_{\lambda  b}{}^{ a}{}_{ c_1} }  \rt)  \xi^{ c_1} 
+ \lt( - {1\over 2} \underline{  \gamma^{ a}_{ c_1  c_2 ,  b } }  
+ \underline{  \gamma^{ a}_{  c_1  d }  \gamma^{ d}_{  b  c_2 }  } 
- \underline{  \gamma^{ a}_{  c_1  d }  \omg_{\lambda  b}{}^{ d}{}_{ c_2 } }   \rt)
 \xi^{ c_1  c_2}  \rt. \cr
&&
\lt. + \lt( - {1\over 3!} \underline{ \gamma^{ a}_{ c_1  c_2  c_3 ,  b } } 
+ {1\over 2} \underline{ \gamma^{ a}_{  c_1  c_2  d }  \gamma^{ d}_{  b  c_3 }  } 
- {1\over 2} \underline{ \gamma^{ a}_{  c_1 c_2  d }  \omg_{\lambda  b}{}^{ d}{}_{ c_3} } \rt)
 \xi^{ c_1  c_2  c_3} \rt] + O( \xi^4) ~, 
\label{k-exp-beta}
\\
&& \cr
 e^{( a)}{}_{ b} 
&=&  \underline{ e^{( a)}{}_{ b} } + \underline{  e^{( a)}{}_{ b,  c} }  \xi^{ c} 
+ {1\over 2} \lt( \underline{  e^{( a)}{}_{ b ,  c_1  c_2 } } - \underline{  e^{( a)}{}_{ b,  d}   \gamma^{ d}_{ c_1  c_2} } \rt)
 \xi^{ c_1  c_2}  \cr
&&
+ \lt( {1\over 3! } \underline{  e^{( a)}{}_{ b ,  c_1  c_2  c_3 } } 
- {1\over 2} \underline{  e^{( a)}{}_{ b ,  d  c_1 } \gamma^{ d}_{ c_2  c_3 } } 
- {1\over 3!} \underline{  e^{( a)}{}_{ b,  d}  \gamma^{ d}_{ c_1  c_2  c_3 } } \rt)  \xi^{ c_1  c_2  c_3} + O( \xi^4)  ~,
\label{e-exp} 
\eea
where we have used eqs.(\ref{z-zhat}, \ref{exp}) and the following result,
\bea
K^{ b}{}_{C, \al} &=& \lt( - \underline{  \gamma^{ b}_{ a  c} } 
+ \underline{  \omg_{\lambda  a}{}^{ b}{}_{ c} } \rt) f^{ a}{}_{,\al} K^{ c}{}_C ~,
\label{K-derivative}
\eea
$\omg_{\lambda a}$ being the spin connection corresponding to $e_{\lambda}^{(\hat a)}{}_b$ as defined in (\ref{e-lambda-def}).

\subsection{Verification up to quadratic order}
\label{as:quadratic}

We substitute the results (\ref{lhs-apriori-B} - \ref{e-exp}) in eq.(\ref{ehat-e}), equate the coefficients at each order up to quadratic order and check if they are satisfied. The fact that the zero-th order terms match simply follow from our construction in \S \ref{ss:method}. The linear term in transverse component of vielbein vanishes, simply because it involves a total covariant derivative of vielbein. The other non-trivial identities obtained are as discussed in \S \ref{s:consistency} which we verify to be true.

\subsection{Verification at cubic order }
\label{as:cubic}

Our goal here is to verify the coefficient of the cubic term in the second equation of (\ref{lhs-fnc}). Given that the computation involved is very tedious, we adopt the following strategy to simplify it. 
\begin{enumerate}
\item
Consider the situation where,
\bea
\del_{\al} \Lambda^{\hat b}{}_{ b} &=& 0~.
\eea
This way one gets rid of all the inhomogeneous terms in the expansions. 

\item
Verify the two terms in the coefficient, namely, 
\bea
\underline{  t_1^{\hat a}{}_{\beta} } = {1\over 6} \Lambda^{\hat a}{}_{ a} f^{ b}{}_{, \beta} 
\underline{  D^{t}_{ c_1}  r^{( a)}{}_{ c_1  c_2  b} }  \xi^{ c_1  c_2  c_3} ~, &&
\underline{  t_2^{\hat a}{}_{\beta} } = {1\over 6} \Lambda^{\hat a}{}_{ a} f^{ b}{}_{, \beta}  
\underline{  r^{( a)}{}_{  c_1  c_2  d}  \omega_{ b}{}^{ d}{}_{ c_3 }}  \xi^{ c_1  c_2  c_3 } ~, 
\label{t1-t2-def}
\cr
&&
\eea
separately by splitting the computation into two different parts. In the first part, to verify $\underline{  t_1^{\hat a}{}_{\beta} }$, one assumes,
\bea
\underline{  \omg_{ a  b  d}} = 0~, \quad \underline{  D_{ c}  \omg_{ a  b  d}} \neq 0 ~, \quad \underline{  D_{ c_1}  D_{ c_2}  \omg_{ a  b  d}} \neq 0 ~,
\label{part-I-cond}
\eea
while to verify $\underline{  t_2^{\hat a}{}_{\beta} }$ one assumes,
\bea
\underline{  \omg_{ a  b  d}} \neq 0~, \quad \underline{  D_{ c}  \omg_{ a  b  d}} = 0 ~, \quad \underline{  D_{ c_1}  D_{ c_2}  \omg_{ a  b  d}} = 0 ~.
\label{part-II-cond}
\eea

\end{enumerate}
With the above strategy in mind, we now follow through the computations for the two parts in order. 


\subsubsection{Part I}
\label{ss:partI}

In this case, the required expressions simplify in the following manner,
\bea
\lambda^{\hat a}{}_{ b} &=&  \Lambda^{\hat a}{}_{ b'} \lt[ \dlt^{ b'}{}_{ b} 
- {1\over 2} \underline{  D_{ c_1}  \omg_{ c_2 ( b)}{}^{( b')} }  \xi^{ c_1  c_2}  
- {1\over 6} \underline{ D_{ c_1}  D_{ c_2}  \omg_{ c_3 ( b)}{}^{( b')} }  \xi^{ c_1  c_2  c_3} \rt] + O( \xi^4)  ~,  
\label{lambda-I} \\
&& \cr
k^{ a}{}_{\beta} &=& f^{ b}{}_{, \beta} \lt[ \dlt^{ a}{}_{ b} 
- \underline{  \gamma^{ a}_{  b  c}}   \xi^{ c} 
+ ( - {1\over 2} \underline{  \gamma^{ a}_{  c_1  c_2 ,  b } } 
+ \underline{  \gamma^{ a}_{  c_1  d }  \gamma^{ d}_{  b  c_2 } } )  \xi^{ c_1  c_2} 
\rt. \cr
&& \lt. + ( - {1\over 6} \underline{  \gamma^{  a}_{  c_1  c_2  c_3 ,  b } }  
+ {1\over 2} \underline{  \gamma^{ a}_{  c_1  c_2  d }  \gamma^{ d}_{  b  c_3 } } ) 
 \xi^{ c_1  c_2  c_3} \rt]  + O( \xi^4) ~, 
\label{k-I} \\
 e^{( a)}{}_{ b}  &=& \underline{ e^{( a)}{}_{ e}} \lt[ \dlt^{ e}{}_{ b} 
+ \underline{ \gamma^{ e}_{ b  c}}  \xi^{ c} 
+ {1\over 2} ( \underline{ \gamma^{ e}_{ b  c_2,  c_1} } 
+ \underline{ \gamma^{ e}_{ d  c_1}  \gamma^{ d}_{ b  c_2} }  
- \underline{ \gamma^{ e}_{ b  d}  \gamma^{ d}_{  c_1  c_2} }
- \underline{ D_{ c_1}  \omg_{ c_2}{}^{ e}{}_{ b} } )   \xi^{ c_1  c_2}  \rt. \cr
&&
+ {1\over 6} \lt\{ \underline{  \gamma^{ e}_{ c_1  b,  c_2  c_3 } } 
+ 2 \underline{ \gamma^{ e}_{ c_1  d}  \gamma^{ d}_{ c_2  b,  c_3 } }    
- 3 \underline{  \gamma^{ e}_{ d  b,  c_1 }   \gamma^{ d}_{ c_2  c_3 } }
+ \underline{ \gamma^{ e}_{ d  c_2 ,  c_3 }  \gamma^{ d}_{ c_1  b} }  
- \underline{ \gamma^{ e}_{ b  d }  \gamma^{ d}_{ c_1  c_2  c_3 } } 
- 3 \underline{  \gamma^{ e}_{ c_1  f}  \gamma^{ f}_{ b  d }  \gamma^{ d}_{ c_2  c_3 } } \rt. \cr
&& 
+ \underline{ \gamma^{ e}_{ c_1  d}  \gamma^{ d}_{ c_2  f}  \gamma^{ f}_{ c_3  b} }   
- \underline{ D_{ c_1}  D_{ c_2}  \omg_{ c_3}{}^{ e}{}_{ b} } 
- 3 \underline{ \gamma^{ f}_{ c_1  b}  D_{ c_2}  \omg_{ c_3}{}^{ e}{}_{ f} }  \cr
&& \lt. \lt. 
+ \underline{ \gamma^f_{c_1 c_2} e_{(a')}{}^e \lt( D_f  D_{c_3} - D_{c_3} D_f \rt) e^{(a')}{}_b } \rt\} \xi^{c_1 c_2 c_3} \rt]  + O( \xi^4) ~, 
\label{e-I}
\cr &&
\eea
We substitute the results (\ref{lambda-I}, \ref{k-I}, \ref{e-I}) into RHS of eq.(\ref{ehat-e}) and extract the contribution at cubic order. At this stage all terms involving (covariant derivatives of) spin connection cancel out, as we would require them to. After a further simplification and equating the result to the first expression in (\ref{t1-t2-def}) one arrives at the following identity,
\bea
D_{c_1} r^{a}{}_{c_2 c_3 b} \xi^{ c_1 c_2 c_3} &=& 
\lt[ \gamma^a_{c_1 b, c_2 c_3} - \gamma^a_{c_1 c_2 c_3 , b } 
+ 2 \gamma^a_{c_1 d} \gamma^d_{c_2 b, c_3 }    
- 3 \gamma^a_{db, c_1 }  \gamma^{d}_{c_2 c_3 } 
+ \gamma^a_{d c_2 , c_3 } \gamma^d_{c_1 b}  
-3 \gamma^a_{f c_2, c_1} \gamma^f_{ b c_3} \rt. \cr
&&
- 3 \gamma^a_{ f c_3 } \gamma^f_{ c_1 c_2 , b }  
- 3 \gamma^a_{c_1 f} \gamma^f_{b d } \gamma^{d}_{c_2 c_3 }  
+4 \gamma^a_{d c_1} \gamma^d_{f c_2} \gamma^f_{ b c_3}  
+ 3 \gamma^a_{fd} \gamma^{d}_{c_1 c_2} \gamma^f_{ b c_3}  \cr
&& \lt. 
- \gamma^a_{b d } \gamma^{d}_{c_1 c_2 c_3 } 
+ 3 \gamma^a_{ c_1 c_2 d } \gamma^{d}_{ b c_3 }  
- \gamma^f_{c_1 c_2} r^a{}_{b f c_3}    \rt] \xi^{c_1 c_2 c_3}  ~.
\label{D-r-identity}
\eea
The above identity can indeed be shown to be true. 

Notice that the last term in the square bracket of the above equation, namely $- \gamma^f_{c_1 c_2} r^a{}_{b f c_3}$ originates from the last term on the RHS of (\ref{e-I}). This is due to the use of the following identity,
\bea
(D_{ c} D_{ d} - D_{d} D_{c}) e^{(a)}{}_{b} &=& - r^{f}{}_{ b c d} e^{(a)}{}_{ f} ~.
\label{D-comm-e}
\eea
We therefore identify this term as being originated from spin connection terms. The rest of the terms on the RHS of (\ref{D-r-identity}) can be obtained by formally setting all the spin connection terms to zero in the computation to start with. It will be useful for our computation in Part II to identify all these terms together as,
\bea
\underline{ \rho^a{}_{b} } &=& ( \underline{ D_{c_1} r^{a}{}_{c_2 c_3 b} } 
+ \underline{ \gamma^f_{c_1 c_2} r^a{}_{b f c_3} } ) \xi^{ c_1 c_2 c_3} ~.
\label{rho-def}
\eea


\subsubsection{Part II}
\label{ss:partII}

In this case, we first impose the conditions (\ref{part-II-cond}) on (\ref{lambda-exp}, \ref{k-exp-beta}, \ref{e-exp}), substitute the results in the RHS of (\ref{ehat-e}) and compute the contribution at cubic order. All the terms that appear in the final expression can be divided into two parts: (1) ones that originate from spin connection terms and (2) ones that do not. The latter is same as that appear in the previous computation and therefore is given by eq.(\ref{rho-def}). Below we shall focus on computing the spin connection terms, with the condition (\ref{part-II-cond}) imposed of course.

The relevant parts of the expressions in (\ref{lambda-exp}, \ref{k-exp-beta}, \ref{e-exp}) are as follows,
\bea
\lambda^{\hat a}{}_b &=& \Lambda^{\hat a}{}_{ b'} \lt[ \dlt^{ b'}{}_{ b} - \underline{ \omega_{ c_1( b)}{}^{( b')}}  \xi^{ c_1}
+ {1\over 2} \underline{  \omg_{c_1 (b)}{}^{(e)} \omega_{c_2 ( e)}{}^{( b')}  }  \xi^{ c_1  c_2}  
- {1\over 6} \underline{ \omg_{c_1 (b)}{}^{(e)} \omg_{c_2 (e)}{}^{(f)} \omega_{ c_3 ( f)}{}^{( b')} }  \xi^{ c_1  c_2  c_3} 
\rt] ~, \cr
k^{ a}{}_{\beta} &\to& f^{ b}{}_{, \beta} \lt[ \dlt^{ a}{}_{ b} 
+ \lt( - \underline{ \gamma^{ a}_{ b  b_1} } 
+ \underline{ \omg_{b}{}^{ a}{}_{ b_1} }  \rt)  \xi^{ b_1} 
+ \lt( - {1\over 2} \underline{  \gamma^{ a}_{ b_1  b_2 ,  b } }  
+ \underline{  \gamma^{ a}_{  b_1  d }  \gamma^{ d}_{  b  b_2 }  } 
- \underline{  \gamma^{ a}_{  b_1  d }  \omg_{b}{}^{ d}{}_{ b_2 } }   \rt)
 \xi^{ b_1  b_2}  \rt. \cr
&&
\lt. 
- {1\over 2} \underline{ \gamma^{ a}_{  b_1  b_2  d }  \omg_{b}{}^{ d}{}_{ b_3} }  \xi^{ b_1  b_2  b_3} \rt] + O( \xi^4) ~, \\
e^{(a)}{}_b &\to& \underline{e^{(a)}{}_e } \lt[ \dlt^e{}_b + \lt\{ \underline{ \gamma^e_{bc} } - \underline{ \omg_c{}^e{}_b } \rt\} \xi^{c} \rt. \cr
&&
+ {1\over 2} \lt\{ \underline{ \gamma^e_{b c_2, c_1 } } 
+ \underline{ \gamma^d_{b c_2} \gamma^e_{c_1 d} } 
- \underline{ \gamma^{d}_{c_1 c_2} \gamma^e_{bd} }
- 2 \underline{ \gamma^d_{c_1 b} \omg_{c_2}{}^e{}_d }
+ \underline{ \omg_{c_1}{}^e{}_f \omg_{c_2}{}^f{}_b } \rt\}
\xi^{c_1 c_2}  \cr
&&
+ {1\over 6} \lt\{ 
- 3 \underline{ \gamma^d_{c_1 b, c_2} \omg_{c_3}{}^e{}_d  } 
- 3 \underline{ \gamma^d_{c_1 b}  \gamma^f_{c_2 d} \omg_{c_3}{}^e{}_f }
+ 3 \underline{ \gamma^f_{db} \gamma^{d}_{c_2 c_3 } \omg_{c_1}{}^e{}_f  }
+ 3 \underline{ \gamma^d_{c_1 b} \omg_{c_3}{}^e{}_f \omg_{c_2}{}^f{}_d }
\rt. \cr
&& \lt. \lt. 
- \underline{ \omg_{c_3}{}^e{}_f \omg_{c_1}{}^f{}_d \omg_{c_2}{}^d{}_b } 
- \underline{ \gamma^d_{c_1 c_2}} (\underline{\omg_{c_3}{}^e{}_f \omg_d{}^f{}_b } - c_3 \leftrightarrow d)
\rt\}
\xi^{c_1 c_2 c_3}  \rt] ~,
\label{e-II}
\eea
Notice that, in addition to imposing (\ref{part-II-cond}), we have removed terms that will necessarily contribute $\omg$-independent terms at cubic order on the RHS of (\ref{ehat-e}). Upon substituting the above expressions in the RHS 
of (\ref{ehat-e}) and keeping only $\omg$-dependent terms one one can show that all $\omg^3$ and most of the $\omg^2$ terms cancel out giving rise to the following expression,
\bea
&& {1\over 6}  \Lambda^{\hat a}{}_{a}  f^b{}_{,\beta} 
\lt[ \underline{ r^{(a)}{}_{c_1 c_2 d} \omg_{b}{}^{ d}{}_{ c_3 } }  
- \underline{ \gamma^{d}_{c_1 c_2}  e^{(c)}{}_b }  ( \underline{ \omg_{c_3}{}^{(a)}{}_{(e)} \omg_{d}{}^{(e)}{}_{(c)} } 
- c_3 \leftrightarrow d  )     \rt]     \xi^{ c_1 c_2 c_3} ~,
\label{part-II-result}
\eea
where the second term in the square bracket originates from the last term on the RHS of (\ref{e-II}). This can be further manipulated in the following way,
\bea
&& - \underline{ \gamma^{d}_{c_1 c_2}  e^{(c)}{}_b }  ( \underline{ \omg_{c_3}{}^{(a)}{}_{(e)} \omg_{d}{}^{(e)}{}_{(c)} }
- c_3 \leftrightarrow d  )   ~, \cr
&=& \underline{ \gamma^{d}_{c_1 c_2}  e^{(c)}{}_b }  ( \underline{ D^{t}_{c_3} \omg_d{}^{(a)}{}_{(c)} }  
- \underline{ \omg_{c_3}{}^{(a)}{}_{(e)} \omg_{d}{}^{(e)}{}_{(c)} } - c_3 \leftrightarrow d  ) ~, \cr
&=& - \underline{ \gamma^f_{c_1 c_2} r^{(a)}{}_{b f c_3 } } ~,
\eea
where in the second line we have used $\underline{ D^{t}_{c_3} \omg_d{}^{(a)}{}_{(c)} } = 0$, which follows from condition 
(\ref{part-II-cond}). Therefore, while the first term in (\ref{part-II-result}) is precisely $\underline{ t^{\hat a}_{2 \beta} } $ in (\ref{t1-t2-def}), the second term, combined with the $\omg$-independent terms, i.e. $\rho^a{}_b$ in (\ref{rho-def}), gives rise to $\underline{ t^{\hat a}_{1 \beta} } $.

\end{document}